\RequirePackage{fix-cm}
\documentclass[smallextended]{svjour3}       
\smartqed  
\usepackage{graphicx}
\usepackage{amssymb} 

\usepackage{subfigure}
\usepackage{amsmath}
\usepackage{amsfonts}
\usepackage{amsbsy}
\usepackage{times}
\usepackage{mathptmx}
\usepackage[left]{lineno}

 \journalname{J. Braz. Soc. Mech. Sci. Eng.}
\begin{document}

\title{Transitions between smooth and rough surfaces in turbulent channel flows for d- and k-type rough elements\thanks{This is a pre-print of an article published in Journal of the Brazilian Society of Mechanical Sciences and Engineering, 40:187, 2018. The final authenticated version is available online at: http://dx.doi.org/10.1007/s40430-018-1113-9}
}


\author{Gabriel Maltese Meletti de Oliveira \and
        Erick de Moraes Franklin 
}


\institute{Gabriel Maltese Meletti de Oliveira \at
              School of Mechanical Engineering, University of Campinas - UNICAMP \\
							 \email{gabrielmeletti@gmail.com}               
									 \and
           Erick de Moraes Franklin (corresponding author)\at
              School of Mechanical Engineering, University of Campinas - UNICAMP \\
              Tel.: +55-19-35213375\\							
              orcid.org/0000-0003-2754-596X\\
              \email{franklin@fem.unicamp.br}             
}

\date{Received: date / Accepted: date}

\maketitle

\begin{abstract}
\begin{sloppypar}
This paper presents an experimental study on the transition from smooth to rough walls, and back to the smooth one, in turbulent closed-conduit flows. These transitions cause a shift on flow velocity profiles that changes their parameters when compared to the flow over a smooth wall. Different water flow rates were imposed in a closed conduit of rectangular cross section, where rough elements consisting of cavities of d- and k-type were positioned covering a part of the bottom wall of the test section. Reynolds numbers based on the channel half-height were moderate, varying between 7800 and 9600, and the regime upstream of the rough elements was hydraulically smooth. Experimental data for this specific case remain scarce and the involved physics rests to be understood. The flow field was measured by low frequency PIV (particle image velocimetry) and by flow visualization, the latter using a continuous 0.1 W laser, a high-speed camera, and scripts written by the authors. From the instantaneous fields measured with PIV, the mean velocities, fluctuations, shear stresses and turbulence production were computed. The results show the presence of oscillations in Reynolds stress and turbulence production, that are higher for the k-type roughness and were not shown in previous experimental works. From the high-speed movies, the angular velocities and frequencies of vortices in the cavities were computed, and the occurrence of fluid ejection from the cavities to upper layers of the flow was observed. A relation between the angular velocities of inner-cavities vortices and the oscillations in Reynolds stress and turbulence production is proposed.
\end{sloppypar}
\end{abstract}

\section*{Nomenclature}
\label{section:nomenclature}

\subsection*{Roman symbols}
\begin{tabular}{l l}
$B$ & constant\\
$d$ & diameter (m)\\
$d_h$ & hydraulic diameter (m)\\
$f$ & Darcy friction factor\\
$f$ & frequency (Hz)\\
$h$ & channel half height (m)\\
$H$ & channel height (m)\\
$k$ & height of rough elements (m)\\
$L$ & wavelength (m)\\
$l_v$ & viscous length (m)\\
$P$ & turbulence production term (m$^2$/s$^3$)\\
$Q$ & volumetric flow rate (m$^3$/h)\\
$Re$ & Reynolds number based on the boundary-layer thickness\\
$Re_{dh}$ & Reynolds number based on the hydraulic diameter\\
$t$ & time (s)\\
$u$ & longitudinal component of the velocity (m/s)\\
$u_*$ & shear velocity (m/s)\\
$v$ & vertical component of the velocity (m/s)\\
$v_t$ & tangential velocity (m/s)\\
$w$ & distance between rough elements (m)\\
$x$ & horizontal coordinate (m)\\
$y$ & vertical coordinate (m)\\
\end{tabular}

\subsection*{Greek symbols}
\begin{tabular}{l l}
$\delta$ & boundary-layer thickness (m)\\
$\Delta$ & displacement\\
$\kappa$ & von K\'arm\'an constant\\
$\lambda$ & ratio between the projected frontal and parallel areas\\
$\mu$ & dynamic viscosity (Pa.s)\\
$\nu$ & kinematic viscosity (m$^2$/s)\\
$\Omega$ & angular velocity (rad/s)\\
$\rho$ & specific mass (kg/m$^3$)\\
$\tau$ & shear stress (N/m$^2$)\\
\end{tabular}

\subsection*{Subscripts}
\begin{tabular}{l l}
$0$ & relative to the smooth surface\\
$0$ & relative to the origin of the vertical coordinate\\
$bla$ & relative to Blasius correlation\\
$d$ & relative to the displaced coordinate system\\
$dh$ & relative to the hydraulic diameter\\
$s$ & sand equivalent\\
$v$ & relative to the viscous layer\\
\end{tabular}

\subsection*{Superscripts}
\begin{tabular}{l l}
$+$  & normalized by the viscous length $l_v$ or by the shear velocity $u_*$\\
$\overline{\quad}$ & averaged in time\\
$'$ & fluctuation\\ 		
\end{tabular}

\section{Introduction}
 
Turbulent flows over smooth and rough walls are encountered in both nature and industry. In many cases, there are transitions from walls with small rough elements to walls with larger rough elements, and vice versa. One typical example is the transition that the wind suffers when it blows over a surface that changes abruptly from a relatively smooth boundary, such as the ocean and flat lands, to a rougher one, such as forests and cities. In industry, it is common that pipes change their internal surface from a smooth steel wall to a rough corrugated wall, such as the flexible pipes employed in petroleum pipelines.

A turbulent boundary layer over a flat wall has different regions: an external layer, far from the wall, where the flow is dominated by inertia and pressure gradients and the length scale is the boundary-layer thickness $\delta$; a viscous layer, that exists very close to the wall if the surface is smooth enough, where the flow is dominated by viscous effects and the length scale is the viscous length $l_v = \nu/u_*$, $\nu$ being the kinematic viscosity and $u_*$ the shear velocity; an overlap layer between the external and viscous ones, whose velocity profile is logarithmic; and a buffer layer, matching the viscous and overlap layers. The turbulent boundary layer is in hydraulic smooth regime when the rough elements of the wall are smaller than some units of the viscous length, and it is in the transition or rough regimes otherwise.

The transition from a smooth to a rough wall, and from a rough to a smooth wall, changes the structure of the flow. In the case of fully-developed boundary layers, such as channel and pipe flows, it is common to describe the effect of roughness based on the deviations of mean velocities from the smooth wall case. The first studies on the effects caused by rough walls on turbulent flows were probably the ones made by Hagen \cite{Hagen} and Darcy \cite{Darcy}, in the case of channel flows in fully rough regime. The concept of smooth and rough regimes was not established at the time Hagen and Darcy did their work; therefore, they did not describe the flow in terms of deviation from a smooth wall, and were mainly interested in pressure losses caused by rough walls. Later, Nikuradse \cite{Nikuradse} performed exhaustive experiments with rough pipes in order to determine the effects of closed spaced rough elements, for which he employed sand, on turbulent pipe flows. Nikuradse compared the flows in rough and smooth pipes, observed that the flows in hydraulic rough and transitional regimes were deviations from the smooth regime, and proposed to describe the velocity profiles of transitional and rough regimes by subtracting a deviation function from the smooth profile. Disregarding the external layer,

\begin{equation}
	u^+ = \frac{1}{\kappa} \ln y^+ + B - \Delta B
	\label{nikuradse_eq}
\end{equation}

\noindent where $u^+ = \overline{u}/u_*$, $\overline{u}$ is the flow mean velocity, $y^+ = y/l_v$, $\kappa = 0.41$ is the von K\'arm\'an constant, $B = 5.0$ is the constant corresponding to hydraulic smooth profiles, and $\Delta B$ is the deviation or roughness function. Without the last term on its RHS, Eq. \ref{nikuradse_eq} is valid for the smooth regime. For the developed rough regime, Nikuradse \cite{Nikuradse} found

\begin{equation}
	\Delta B = \frac{1}{\kappa} \ln k_s^+ - 3.5
	\label{deviation_function}
\end{equation}

\noindent where $k_s^+ = k_s/l_v$ and $k_s$ is the mean size of the grains of sand on the tube internal surface. Inserting Eq. \ref{deviation_function} in Eq. \ref{nikuradse_eq}, we obtain an equation frequently used for developed rough regimes:

\begin{equation}
	u^+ = \frac{1}{\kappa} \ln \left( \frac{y}{y_0} \right)
	\label{nikuradse_eq2}
\end{equation}

In Eq. \ref{nikuradse_eq2}, $y_0 = k_s/33$. Because Nikuradse \cite{Nikuradse} presented exhaustive data for rough pipe flows, it is common to relate the actual height of rough elements of a given wall, $k$, to the equivalent one in the Nikuradses'case, $k_s$, which is usually called effective, standard, or sand-like roughness. For that, it is necessary to compare Eq. \ref{nikuradse_eq} applied to the actual and Nikuradse's cases, i.e., using the obtained constants for the actual wall and the values obtained by Nikuradse \cite{Nikuradse}. In principle, this procedure can be applied for uniformly distributed rough elements with regular geometries and typical height much smaller than the pipe diameter or the boundary-layer thickness. In these cases, the height of rough elements and the sand-like roughness are expected to be proportional. However, the behavior is different whether the elements distribution over the wall is dense or sparse.

Schlichting \cite{Schlichting_2} studied the effects of the distribution of regular rough elements over the wall for relatively sparse distributions. He defined the solidity $\lambda$ as the ratio between the projected frontal area of rough elements and the projected area parallel to the wall. He found that $k_s/k \sim \lambda$ for the sparse case. It was proposed the existence of a sparse case for $\lambda \leq 0.15$ , where the standard roughness increases with $\lambda$, and a dense case for $\lambda > 0.15$, where the standard roughness decreases with $\lambda$. The behavior of the dense case can be explained by the sheltering caused by rough elements on each other \cite{Jimenez_1}.

Perry et al. \cite{Perry_1} presented experiments on the development of a turbulent boundary layer over rough walls. The experiments were performed with both zero and adverse pressure gradients, and two different arrangements of rough elements. The authors observed that one of the arrangements, consisting of sparse distributed rough elements, followed the Schlichting-Clauser deviation function \cite{Schlichting_2,Clauser_1,Clauser_2}, with the effective roughness being proportional to the height of rough elements. This arrangement is usually named k-type, because the friction is proportional to the height $k$ of rough elements. For the other arrangement, consisting of densely distributed rough elements, Perry et al. \cite{Perry_1} observed that the effective roughness was not proportional to the height of rough elements. Instead, the effective roughness varied with the thickness of the boundary layer. This arrangement is usually named d-type, once the friction is proportional to the boundary-layer thickness $\delta$.

Many studies have been devoted to turbulent flows over rough walls, and to the transition from smooth to rough regimes \cite{Coleman_4,Raupach,Djenidi,Djenidi2,Sutardi,Leonardi2,Leonardi,Schultz,Lee_B,Lee_C}. Today, we know that the changes caused by rough elements have not an universal behavior and depend strongly on the type of rough elements present on the wall. It is usual to distinguish between two main types of arrangements: the k- and d- types. In general terms, in the k-type the effective roughness is directly proportional to the height of each element, while in the d-type the effective roughness is proportional to the boundary-layer thickness \cite{Jimenez_1}. In the case of square rough elements positioned transverse to the flow, the d-type occurs, usually, when the distance between each rough element is equal or lesser than the element height, and the k-type occurs when the distance between each rough element is greater than the element height \cite{Djenidi,Leonardi}. The common explanation for these differences is based on vortex shedding: in the k-type roughness, vortices with characteristic size given by the element height are frequently shed and diffused into the main flow, whereas in the d-type stable vortices are maintained in the cavities between rough elements, with ejections of fluid into the main flow occurring only occasionally.

\begin{sloppypar}
Djenidi et al. \cite{Djenidi} experimentally investigated boundary layers over smooth and d-type rough walls. The experiments were performed in a 2-m-long water tunnel with a 260 mm $\times$ 260 mm cross section. For the tests with the d-type roughness, a wall with 20 rough elements, 5 mm in height and 5 mm in length each, was positioned at the test section. The mean velocities and fluctuations for flows over smooth and d-type rough walls were measured by a two-component LDA (laser Doppler anemometry) device. In addition, in order to observe the structure of the flow in the cavities, they made flow visualizations in the rough case. Djenidi et al. \cite{Djenidi} observed that the turbulent stresses, specially the shear components, and the normal component of the mean velocity are stronger over the rough wall. The authors also observed that the flow in the cavities has a vortical and quasi-steady structure, with ejections of fluid occurring from time to time. They noted that these ejections intensify momentum transfers in the rough case, which explain the increase in the turbulent stresses and normal mean velocities over the rough wall.
\end{sloppypar}

Sutardi and Ching \cite{Sutardi} presented an experimental study on the perturbation of a turbulent boundary layer in smooth regime by one single rough element. The rough element consisted of a 2D cavity on a flat plate, and three different geometries were investigated: square, semicircular and triangular grooves with length $w$ to height $k$ ratio equal to one. The experiments were performed in a 20-m-long wind tunnel, with a 0.91 m $\times$ 0.91 m cross section, and the flow was measured by hot-wire anemometers. The authors found that the perturbations caused by the square groove are more significant than with the other geometries, that the bursting frequency on the grooved wall is higher than on the smooth wall, and that the $xy$ component of the Reynolds stress increases over the groove and has a large relaxation distance, i.e., it does not reach the smooth value just downstream of the groove.

Leonardi et al. \cite{Leonardi} numerically investigated channel flows with d- and k- type rough walls. The authors performed DNS (direct numerical simulation) for Reynolds numbers based on the channel half height between 2800 and 12000, for ratios between the length $w$ and the height $k$ of cavities within $0.5 \leq w/k \leq 59$, and for ratios between the height of rough elements and the channel half height $h$ within $0.1 \leq k/h \leq 0.2$. Leonardi et al. \cite{Leonardi} found that the behavior of d- and k- type roughness is given by the flow field near the rough wall, and not by the outer scales ($\delta$ or $h$). They proposed that the pressure drag is higher than the frictional drag for the k-type, and that the inverse is true for the d-type.

Recently, Lee \cite{Lee_B} numerically investigated the transition of a turbulent boundary layer from a smooth to a k-type rough surface. The author performed DNS, where the roughness consisted of 2D rods aligned in the transversal direction, periodically distributed in the longitudinal direction, with $w/k = 8$, and positioned over the surface, i.e., with a step change. Lee \cite{Lee_B} showed that there are overshoots in mean velocities and second order moments at each rough element, and that these overshoots are created either within the cavities or at the leading edges of rough elements. 

This paper presents an experimental investigation on the transition from smooth to rough walls, and back to the smooth one, in turbulent closed-conduit flows. The flow upstream of the rough surface is turbulent, fully developed, in hydraulically smooth regime, and under moderate Reynolds numbers (ord($10^4$), where $ord$ stands for order of magnitude). Experimental data for this specific case remain scarce and the involved physics rests to be understood. Water flows were imposed in a transparent closed conduit of rectangular cross section, where two different rough walls were individually positioned at the test section, one of d-type and the other of k-type roughness. The rough elements consisted of cavities machined on the top surface of one of the plates that covered the entire bottom of the channel. The flow field was measured by low frequency PIV (particle image velocimetry) and by flow visualization, the latter using a continuous laser, a high-speed camera and scripts written by the authors. From the instantaneous fields measured with PIV, the mean velocities, fluctuations, shear stresses and turbulence production were computed. The results show the presence of oscillations in Reynolds stress and turbulence production, that are higher for the k-type roughness. These strong oscillations were not reported in previous experimental studies and are shown here for the first time. From the high-speed movies, the angular velocities and frequencies of vortices in the cavities were computed, and the occurrence of fluid ejection from the cavities to upper layers of the flow was observed.  We propose a relation between the angular velocities of inner-cavities vortices and the oscillations in Reynolds stress and turbulence production.

 \section{Experimental device}

\begin{figure}[ht]	
 	\centering
 	\includegraphics[width=0.98\columnwidth]{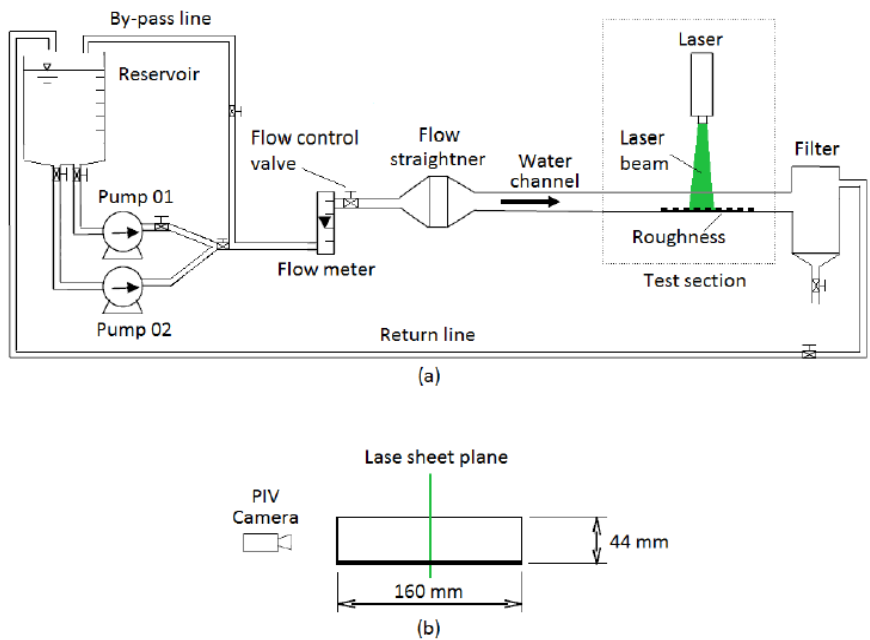}
 	\caption{Layout of the experimental device: (a) lateral view; (b) cross-section.}
 	\label{fig:1}
\end{figure}

The experimental device consisted of a water reservoir, two centrifugal pumps in parallel, an electromagnetic flow meter, a flow straightener, a 5 m long channel of rectangular cross section, a settling tank, and a return line. The channel was a closed conduit of transparent material (plexiglass) with a 160 mm-wide by 50 mm-high cross section. The test section was 1 m long and started at 3 m (40 hydraulic diameters) downstream of the channel inlet, and the remaining 1 m section connected the test section exit to the settling tank. The water flowed in a closed loop following the above order of description. Figure \ref{fig:1} shows the layout of the experimental device.

The centrifugal pumps had their volute and rotor in brass in order to avoid corrosion particles in the water, as these can disturb the PIV (particle image velocimetry) measurements. The flow rates were adjusted by a set of valves. The flow straightener consisted of a divergent-convergent nozzle filled with glass spheres with a mean diameter of $d$ = 3 mm, and its function was to homogenize the water flow.
 
\begin{figure}[ht]	
 	\centering
 	\includegraphics[width=0.8\columnwidth]{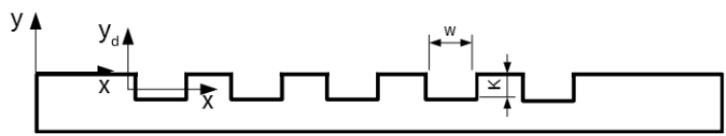}
 	\caption{Geometry of the rough plates and coordinates systems adopted in the present paper.}
 	\label{fig:geometry}
\end{figure}

\begin{figure}[h!]
\begin{center}
	\begin{tabular}{c}
	\includegraphics[width=0.75\columnwidth]{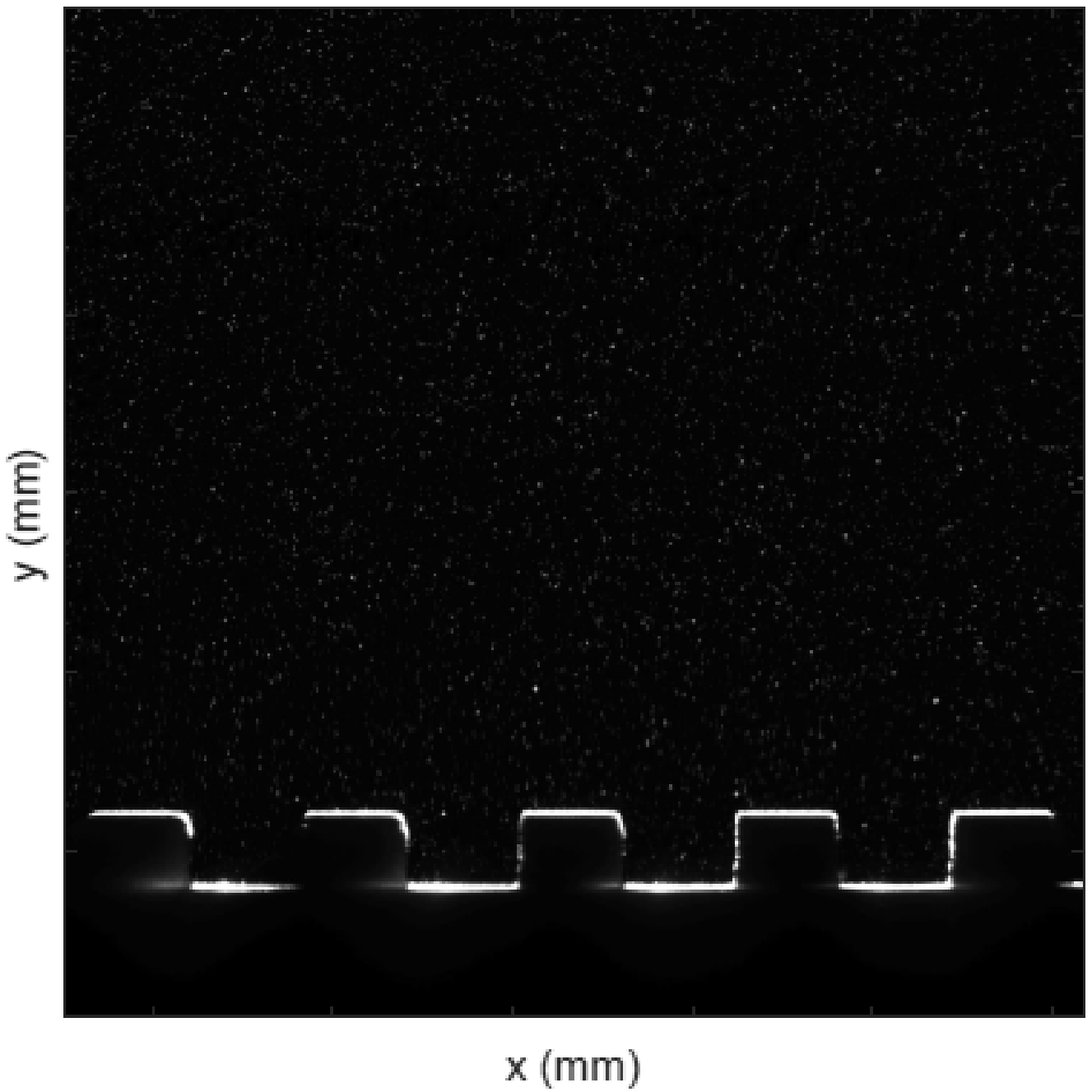}\\
	(a)\\
	\includegraphics[width=0.75\columnwidth]{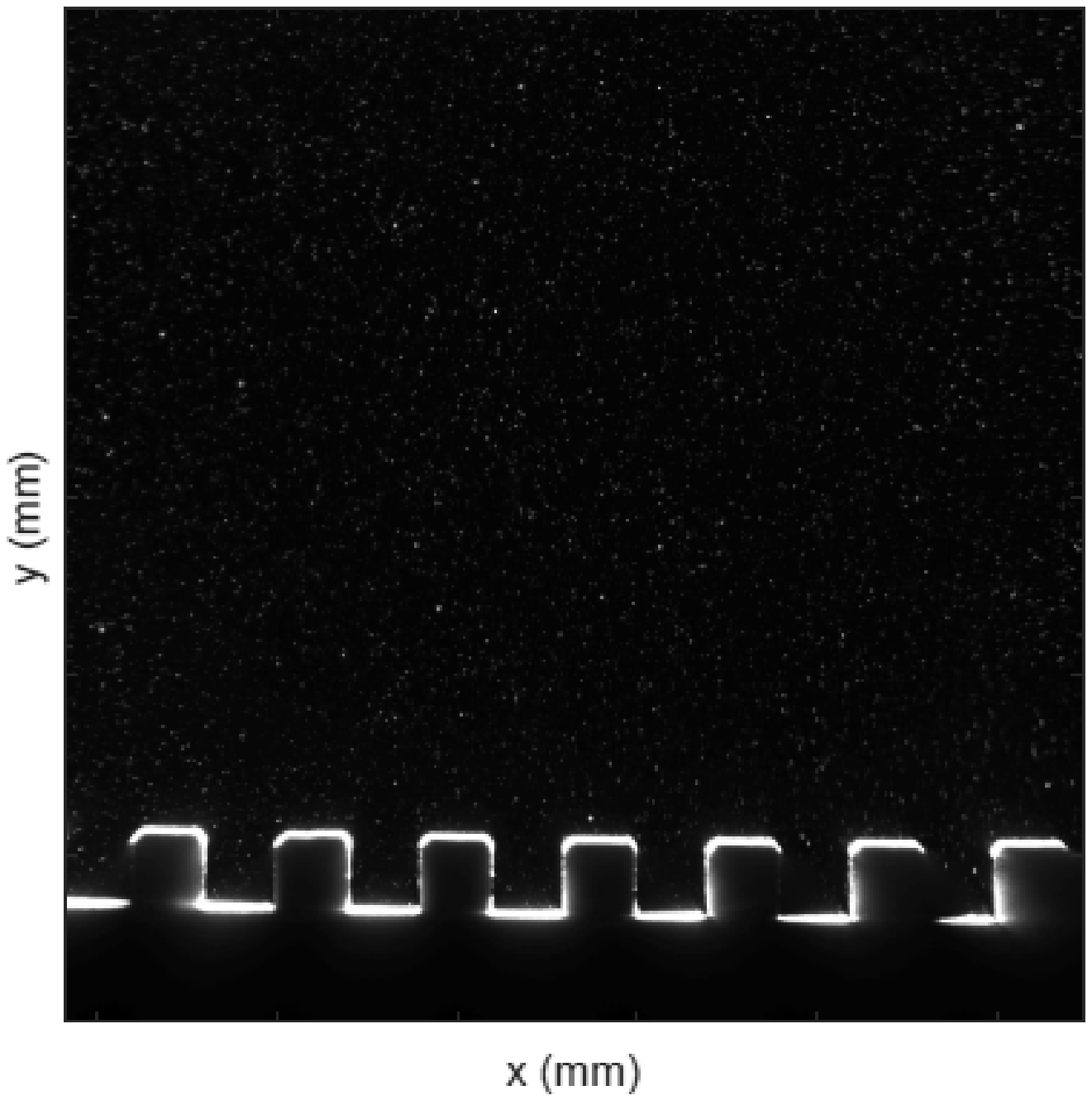}\\
	(b)
	\end{tabular} 
\end{center}
 	\caption{PIV images for the flow over (a) k-type and (b) d-type rough elements.}
 	\label{fig:piv_roughness}
\end{figure}

Flat plates 6 mm thick made of PVC were inserted in the 3 m entrance length and in the 1 m final length of the channel, covering its bottom and reducing its height to 44 mm. A 1 m long plate with smooth and rough regions was inserted in the test section, covering its bottom. The smooth regions of the test section plate were 6 mm thick, and the rough region consisted of cavities hollowed on the plate top surface; therefore, the top of each rough element was at 6 mm from the channel bottom wall. Two different rough plates were used: one, made of aluminum painted in black (electrostatic painting), consisting of one single plate where the rough elements were machined on its top surface; and another one consisting of a rough central section made of black polyoxymethylene, where the rough elements were machined, mounted on flat PVC plates painted in black. Each rough section consisted of 20 square rough elements aligned in a direction transverse to the water flow. The rough elements on the aluminum plate were 3 mm long and 2 mm high, and are referred here as k-type, whereas the rough elements on the polyoxymethylene plate where 2 mm long and 2 mm high and are referred here as d-type. Figure \ref{fig:geometry} shows the variables and the coordinates systems adopted in this work. The two rough sections can be seen in Fig. \ref{fig:piv_roughness}, which shows examples of PIV images. The water temperature was between 22 $^oC$ and 29 $^oC$ in all tests.

Two flow rates were used, 8 and 10 m$^{3}$/h, corresponding to cross-section mean velocities $\left\langle U \right\rangle$ of 0.32 and 0.39 m/s, and to Reynolds numbers $Re = \left\langle U \right\rangle\delta /\nu$ of $7.8\cdot 10^{3}$ and $9.6\cdot 10^{3}$, respectively, where $\delta$ is considered as the half-distance between the surface of the PVC plates and the top wall of the channel. In the test section, the regime was hydraulically smooth upstream of the rough elements in all the cases.

Particle image velocimetry was used to acquire the instantaneous fields of the flow, and all the measurements were made at the vertical symmetry plane of the channel. The employed light source was a dual cavity Nd:YAG Q-Switched laser, capable to emit $2\times 130$ mJ at a maximum pulse rate of 15 Hz with a wavelength of 532 nm. The employed camera was a 7.4 $\mu$m $\times$ 7.4 $\mu$m (px$^{2}$) CCD (charge coupled device) camera with a spatial resolution of 2048 px $\times$ 2048 px, capable to acquire pairs of images at a maximum frequency of 10 Hz. When synchronized with the laser, the employed camera acquires pairs of images at a maximum frequency of 4 Hz. For each roughness type, two fields of view were used: the larger one was 70.4 mm $\times$ 70.4 mm and the other was 28.1 mm $\times$ 28.1 mm, both corresponding to a magnification of approximately 0.1. For the computations, we used cross-correlations, with an interrogation area of 16 px $\times$ 16 px and $50\%$ of overlap for the larger field, and with an interrogation area of 32 px $\times$ 32 px and $50\%$ of overlap for the smaller field. These correspond to 256 interrogation areas and a spatial resolution of 0.27 mm for the larger field, and to 128 interrogation areas and a spatial resolution of 0.22 mm for the smaller field. 

\begin{figure}[ht]
 	\begin{minipage}{0.3\linewidth}
 		\begin{tabular}{c}
 			\includegraphics[width=\linewidth]{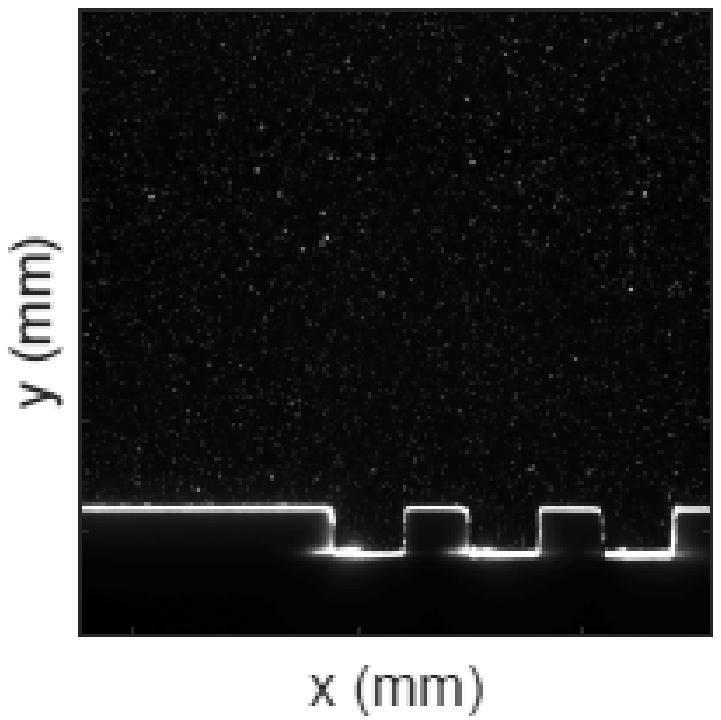}\\
 			(a)
 		\end{tabular}
 	\end{minipage}
 	\hfill
 	\begin{minipage}{0.3\linewidth}
 		\begin{tabular}{c}
 			\includegraphics[width=\linewidth]{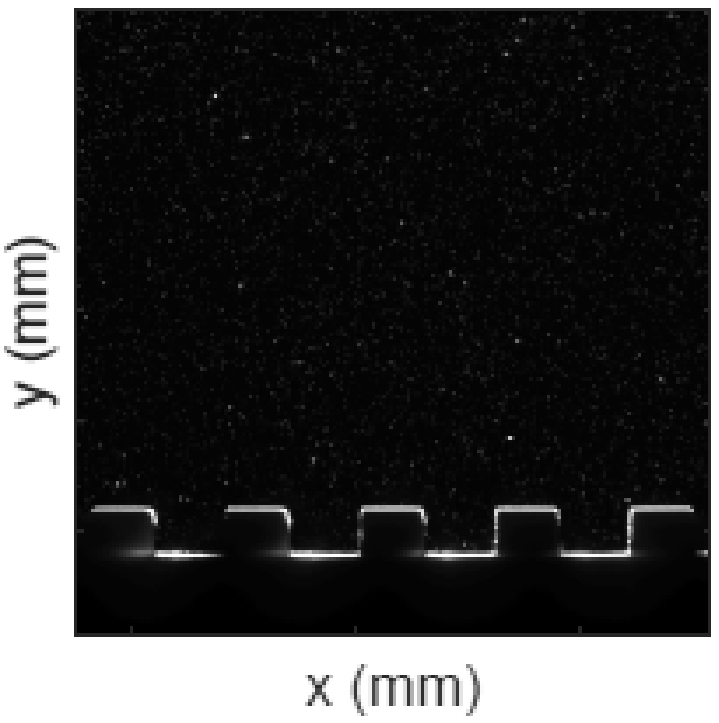}\\
 			(b)
 		\end{tabular}
 	\end{minipage}
	\hfill
 	\begin{minipage}{0.3\linewidth}
 		\begin{tabular}{c}
 			\includegraphics[width=\linewidth]{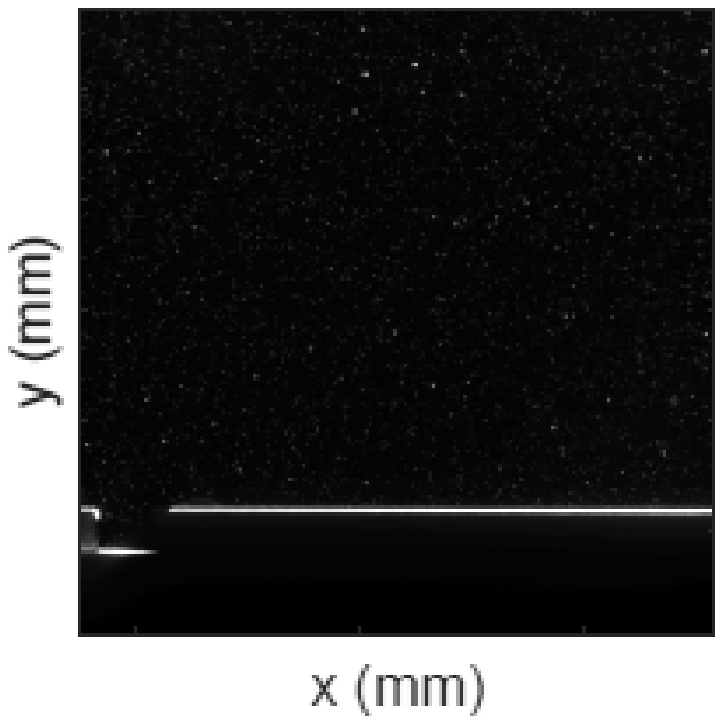}\\
 			(c)
 		\end{tabular}
 	\end{minipage}
 	\caption{Example of different regions imaged by PIV for the k-type roughness: (a) transition from smooth to rough surface; (b) rough surface; (c) transition from rough to smooth surface}
 	\label{fig:montagem_rugosidades}
 \end{figure}

Because the longitudinal dimension of the total field of the camera was smaller than that of the studied region, different test runs had to be made, for both roughness types and flow rates. For the larger field, four test runs were performed for each flow rate: one upstream of the rough elements, one in the transition from the smooth to the rough wall, one in the transition from the rough to the smooth wall, and one downstream of the rough elements. For the smaller field, seven test runs were performed for each flow rate: one upstream of the rough elements, one in the transition from the smooth to the rough wall, four over the rough wall, and one in the transition from the rough to the smooth wall. Figure \ref{fig:montagem_rugosidades} shows an example of different regions imaged by the PIV. Each test run acquired $3000$ image pairs for the larger field and $4000$ image pairs for the smaller field, so that a total of $80000$ image pairs were acquired.

Hollow glass beads with a mean diameter $d$ of 10 $\mu$m and specific gravity of 1.05 were employed as seeding particles. The power of the laser was fixed at $68 \% $ of its maximum power to assure a good balance between the image contrasts and undesirable reflections from the channel walls and plate surfaces. Figure \ref{fig:piv_roughness} shows examples of images acquired by the PIV device under those conditions. Instantaneous velocity fields were computed in fixed Cartesian grids by the PIV controller software. Matlab scripts were written by the authors to post-process these fields in order to compute, for example, the time-averaged velocities, velocity fluctuations, spatio-temporal averaged profiles, shear velocities, stresses, and turbulence production terms.

\begin{figure}[ht]	
 	\centering
 	\includegraphics[width=0.7\columnwidth]{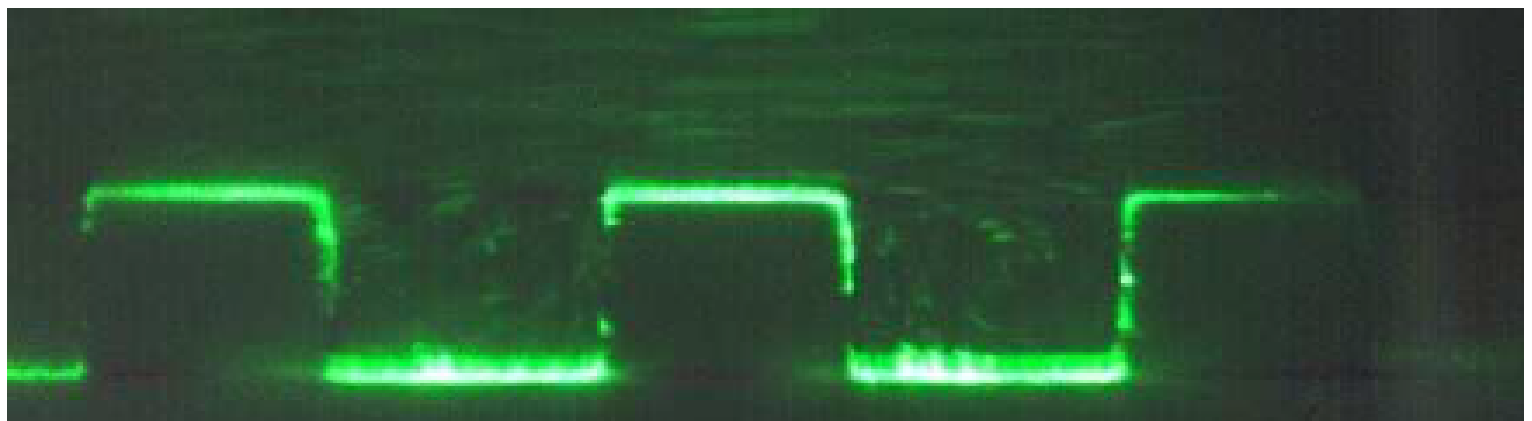}
 	\caption{Flow visualization image obtained for the k-type rough plate. $Re = 9.6\cdot 10^{3}$.}
 	\label{fig:visualization}
\end{figure}

Flow visualization was used to measure the angular velocities of vortices in the cavities. For this, we used a continuous diode laser and a high-speed camera. The laser emitted at a wavelength of 532 nm, and its maximum power was 0.1 W. Hollow glass beads with d = 10 $\mu$m and specific gravity of 1.05 were employed as seeding particles. The camera was of CCD type, with a spatial resolution of 1280 px $\times$ 1024 px, and capable to acquire images at a maximum frequency of 1000 Hz. In order to follow the particles trapped in the cavities, we set the frequency to 300 Hz for the tests with the d-type roughness and to 200 Hz for the tests with the k-type roughness. The images were processed by Matlab scripts written by the authors using a PTV (particle tracking velocimetry) approach. Figure \ref{fig:visualization} shows an example of image obtained for the k-type rough plate.

\section{Results}
 
\subsection{Smooth wall}

The flow upstream of the rough surface was measured for both flow rates. The flow data therefore correspond to a turbulent, fully developed channel flow. After the experiments, the images were cross-correlated by the PIV controller software, and the instantaneous velocity fields were determined. Using Matlab scripts written by the authors, the instantaneous fields were time-averaged, the fluctuation fields were computed and time-averaged, and, because the flow was fully developed, the time-averaged fields were spatially averaged in the longitudinal direction.

\begin{figure}[ht]	
 	\centering
 	\includegraphics[width=0.55\columnwidth]{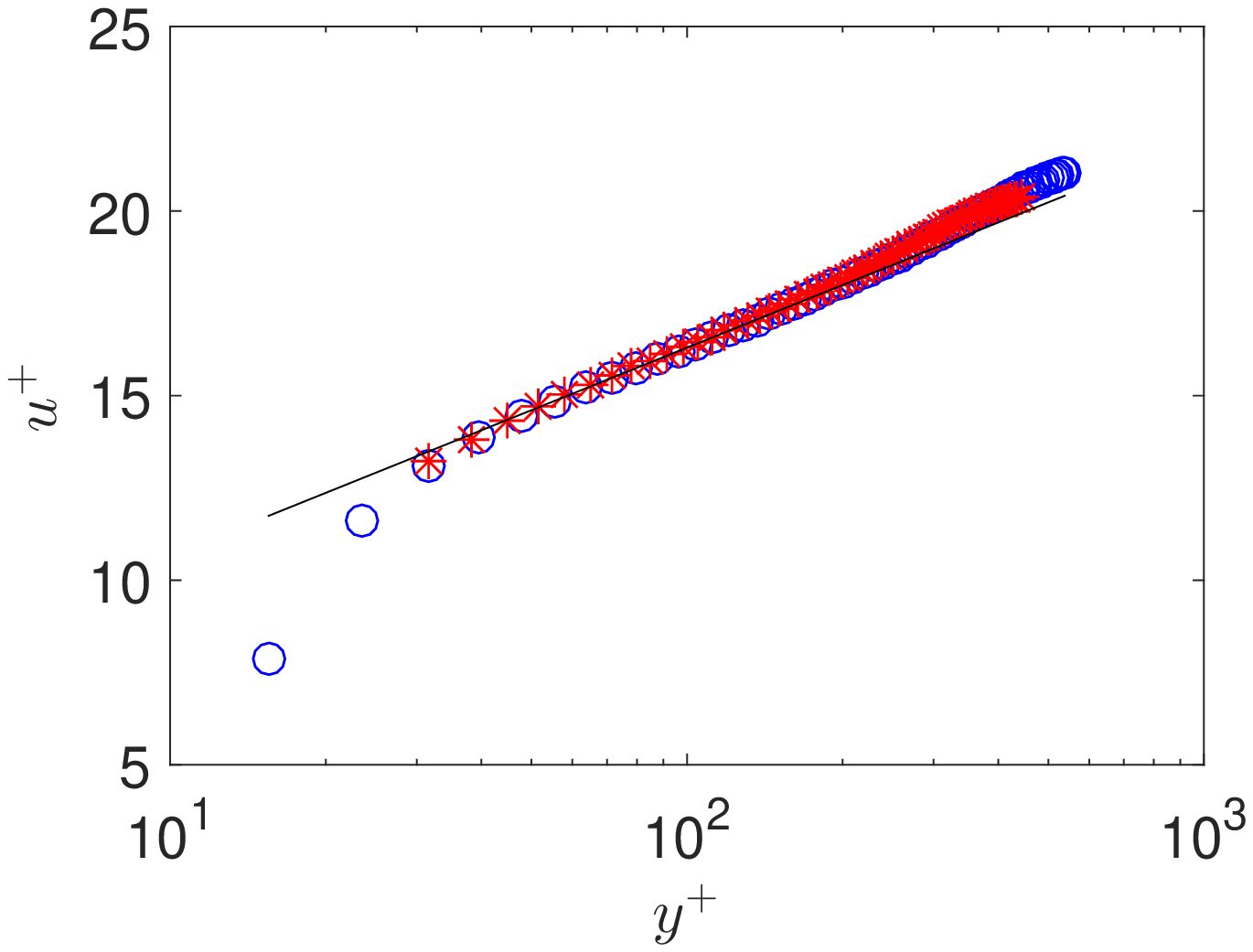}
 	\caption{Mean velocity profiles over the smooth wall. The circles and the asterisks correspond to $Re = 7.8\cdot 10^{3}$ and $Re = 9.6\cdot 10^{3}$, respectively. The line corresponds to Eq. \ref{nikuradse_eq} with $\Delta B =0$.}
 	\label{fig:smooth_log}
\end{figure}

Figure \ref{fig:smooth_log} shows the mean velocity profiles for the two Reynolds numbers. The profiles are plotted in the traditional log-normal scales, with the vertical distance from the wall normalized by the viscous length, $y^+ = yu_{*,0}/ \nu$, and the mean velocity normalized by the shear velocity, $u^+ = u / u_{*,0}$. The circles and the asterisks correspond to $Re = 7.8\cdot 10^{3}$ and $Re = 9.6\cdot 10^{3}$, respectively, and the line corresponds to Eq. \ref{nikuradse_eq} with $\Delta B =0$. The graph was plotted iteratively because the shear velocity $u_{*,0}$ was determined by fitting the experimental data in the logarithmic region ($30 < y^+ < 130$).

Table \ref{tab.1} shows the shear velocity $u_{*,0}$, the law-of-the-wall constant $B_{0}$, the viscous length $l_v$, the experimentally obtained Darcy friction factor $f_{0}$, and the Darcy friction factor from the Blasius correlation $f_{bla} = 0.316 Re_{dh} ^{-1/4}$, where $Re_{dh}$ is the Reynolds Number based on the hydraulic diameter $d_h$ and the cross-section mean velocity $\left\langle U \right\rangle$, in the region upstream of the rough elements, for each flow rate. The measurements showed that the flow in the test section upstream of the rough surface corresponds to a fully-developed turbulent channel flow in hydraulic smooth regime, and the Blasius correlation can be used to estimate flow in this region.

 \begin{table}[ht]
 	\begin{center}
 		\caption{Shear velocity $u_{*,0}$, law-of-the-wall constant $B_{0}$, viscous length $l_v$, experimentally determined Darcy friction factor $f_{0}$ and Darcy friction factor from the Blasius correlation $f_{bla}$ for each water flow rate $Q$, in the region upstream of the rough elements.}
 		\begin{tabular}{c c c c c c c c c c}
 			\hline\hline
 			$Q$ & Type &$\left\langle U \right\rangle$ & $Re_{dh}$ & $B_{0}$ & $u_{*,0}$ &  $l_v$ & $f_{0}$ & $f_{bla}$\\			
			$m^{3}/h$ & $\cdots$ & $m/s$ & $\cdots$ & $\cdots$ & $m/s$ & $m$ & $\cdots$ & $\cdots$\\
 			\hline
 			8 & d & 0.32 & 2.4 $\times$10$^{4}$ & 5.11  & 0.0184 & 4.9 $\times$10$^{-5}$ &  0.0272 & 0.0253\\
 			
 			10 & d & 0.39 & 3.1 $\times$10$^{4}$ & 5.06  & 0.0223 & 4.0 $\times$10$^{-5}$ & 0.0256 & 0.0239\\
			
			8 & k & 0.32 & 2.4 $\times$10$^{4}$& 5.14  & 0.0185 & 4.8 $\times$10$^{-5}$ &  0.0275 & 0.0253\\
 			
 			10 & k & 0.39 & 3.1 $\times$10$^{4}$ & 5.02  & 0.0228 & 3.9 $\times$10$^{-5}$ & 0.0267 & 0.0239\\
 			\hline
 		\end{tabular}
 		\label{tab.1}
 	\end{center}
 \end{table}

\begin{figure}[ht]	
   \begin{minipage}[c]{.49\linewidth}
    \begin{center}
     \includegraphics[width=\linewidth]{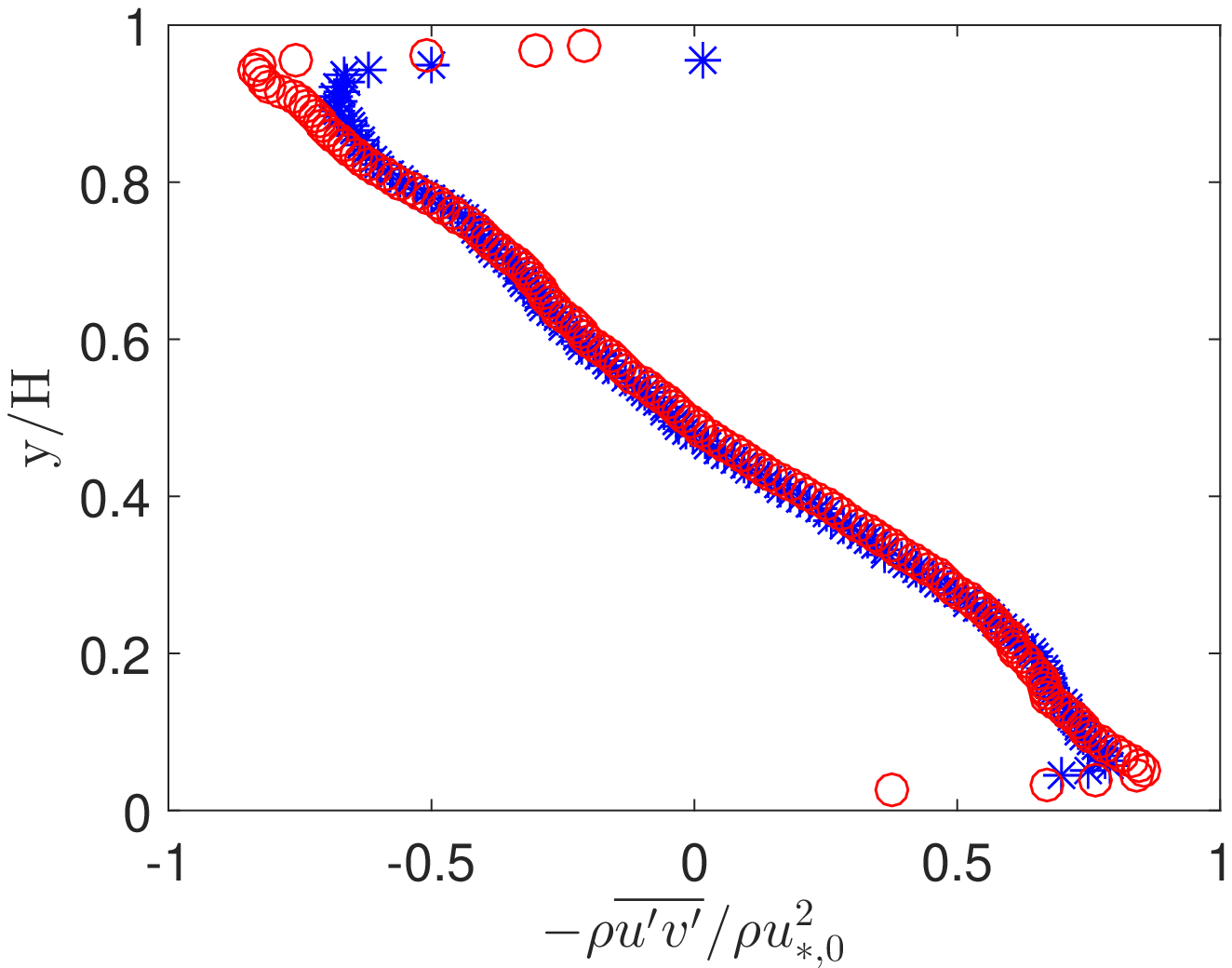}
		(a)
    \end{center}
   \end{minipage} \hfill
   \begin{minipage}[c]{.49\linewidth}
    \begin{center}
      \includegraphics[width=\linewidth]{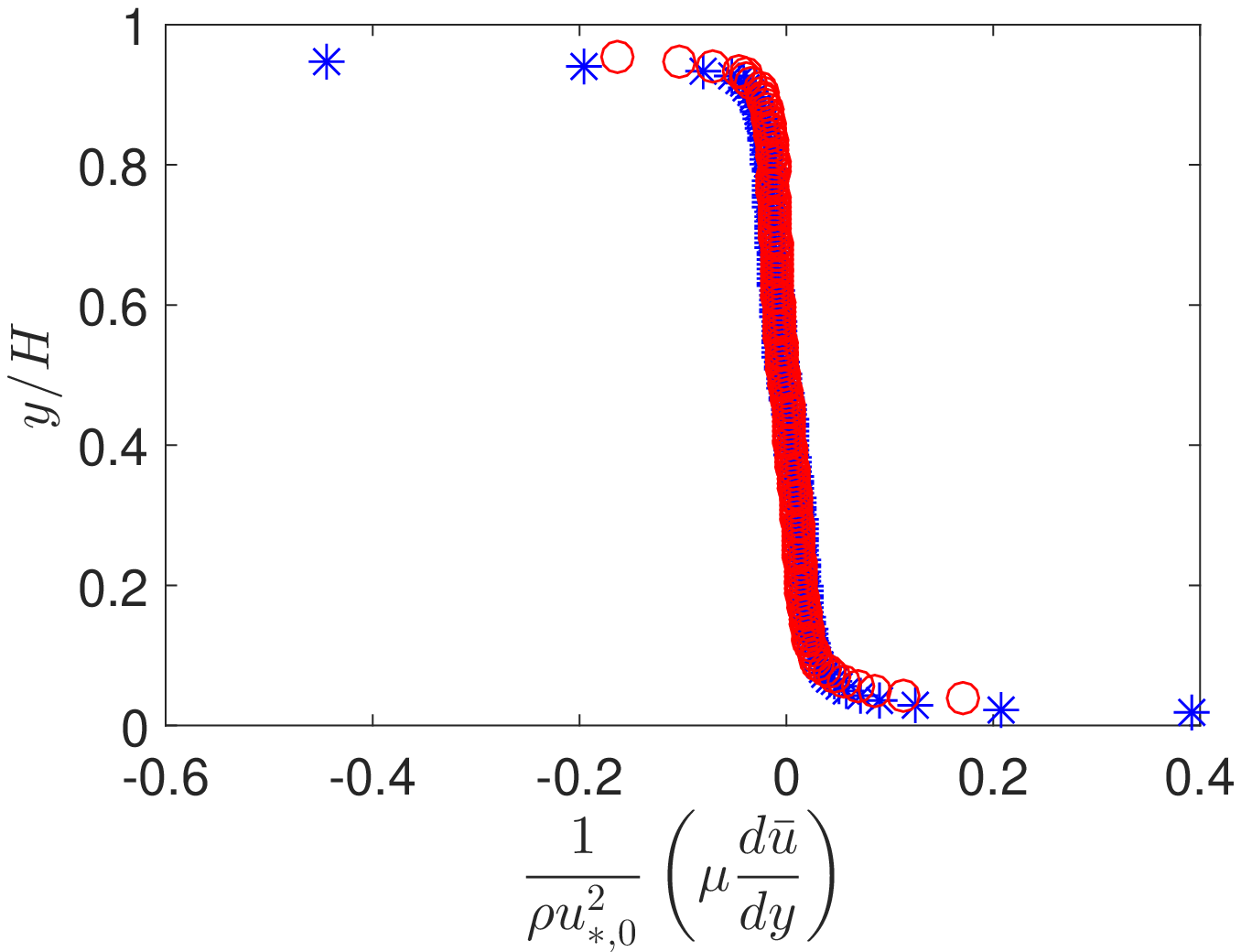}
			(b)
    \end{center}
   \end{minipage}
\caption{(a) Mean profiles of the $xy$ component of the turbulent stress over the smooth wall normalized by $\rho u_{*,0}^2$. (b) Profiles of the $xy$ component of the viscous stress over the smooth wall normalized by $\rho u_{*,0}^2$. The circles and the asterisks correspond to $Re = 7.8\cdot 10^{3}$ and $Re = 9.6\cdot 10^{3}$, respectively.}
\label{fig:stresses_smooth}
\end{figure}

Figure \ref{fig:stresses_smooth}a shows the mean profiles of the $xy$ component of the turbulent stress over the smooth wall in dimensionless form, $y/H$ versus $-\rho \overline{u'v'}/(\rho u_{*,0}^2)$, while Fig. \ref{fig:stresses_smooth}b shows the profiles of the $xy$ component of the viscous stress in dimensionless form, $y/H$ versus $\mu \partial_y \overline{u}/(\rho u_{*,0}^2)$. Two different flow conditions are presented, the circles and the asterisks corresponding to $Re = 7.8\cdot 10^{3}$ and $Re = 9.6\cdot 10^{3}$, respectively. The turbulent and viscous profiles correspond to expected profiles for fully-developed channel flows \cite{Schlichting_1}.

\subsection{Rough walls: spatial analysis}

The flow over the rough elements was measured by PIV for both flow rates and roughness types. Given the relative low frequency of acquisition of image pairs, 4 Hz, and the relative high spatial resolution, 2048 px $\times$ 2048 px, these data are used for an spatial analysis of the flow. As with the flow over the smooth surface, the images were cross-correlated by the PIV controller software, and the instantaneous velocity fields were obtained. Using Matlab scripts written by the authors, the instantaneous fields were then time-averaged, and the fluctuation fields were computed and time-averaged. The time-averaged fields are shown next.

\begin{figure}[ht!]
\begin{minipage}[c]{0.5\textwidth}
\begin{tabular}{c}
\includegraphics[width=\linewidth]{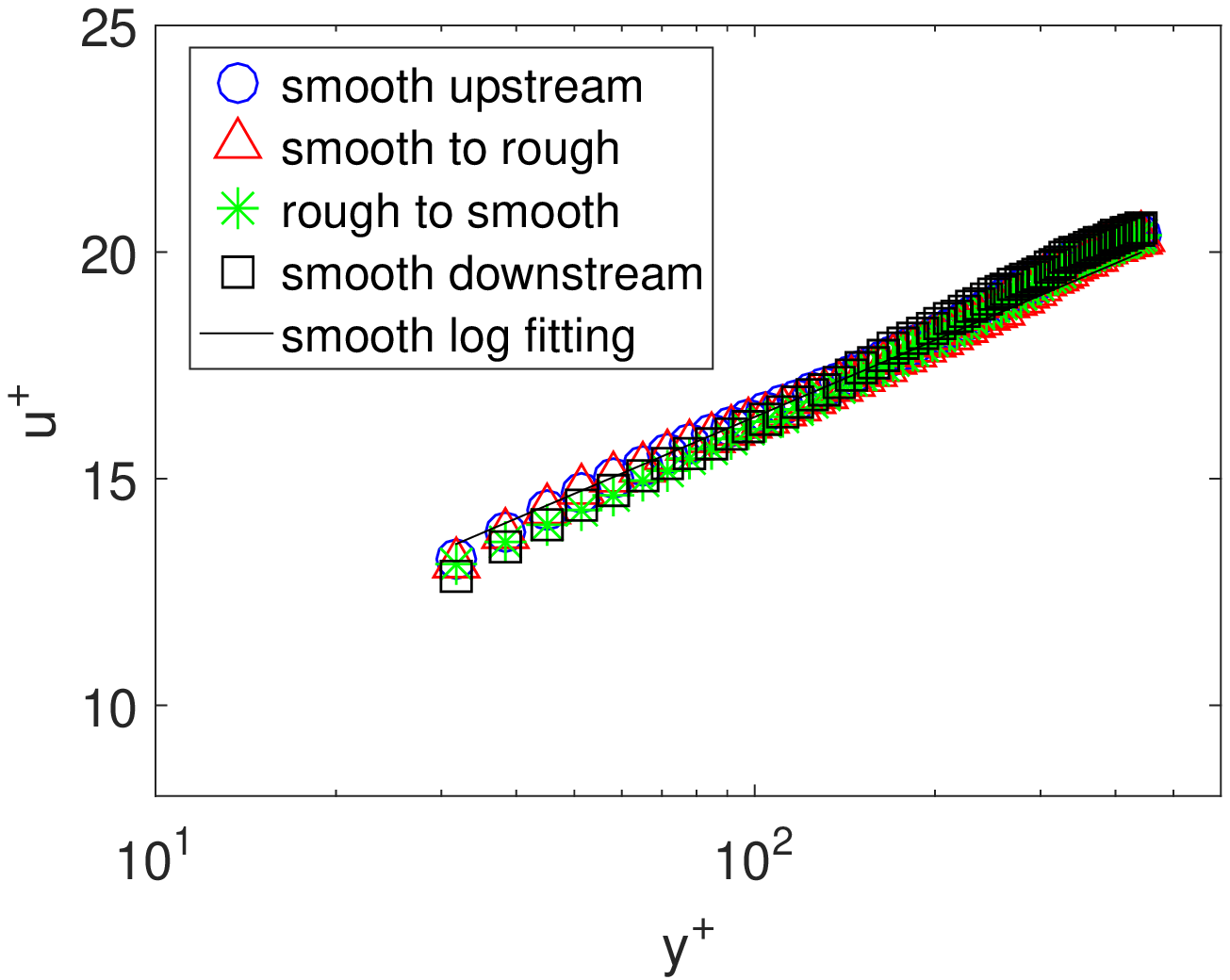}\\
      (a)
\end{tabular}
\end{minipage} \hfill
\begin{minipage}[c]{0.5\textwidth}
\hspace{\fill}
\begin{tabular}{c}
\includegraphics[width=\linewidth]{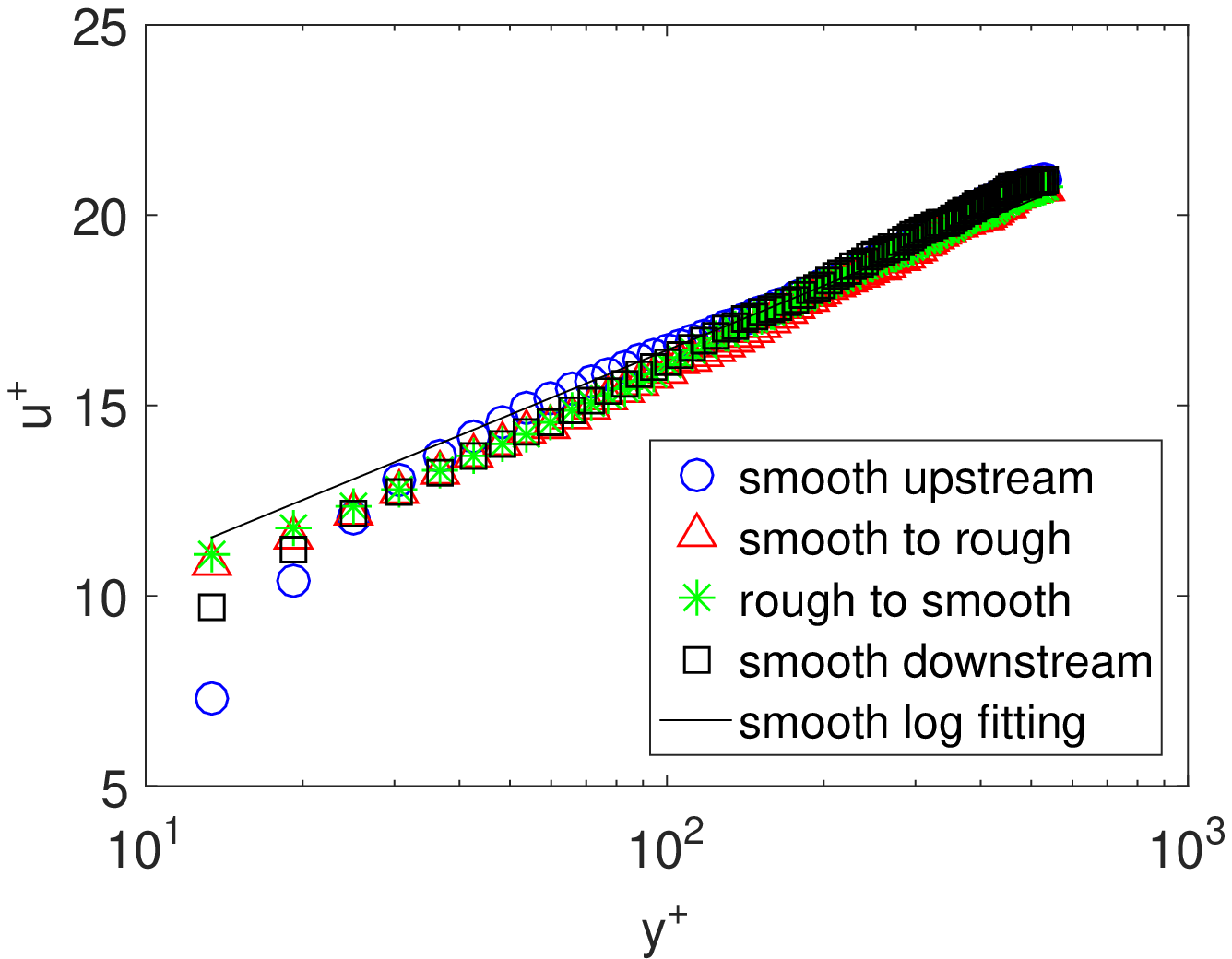}\\
      (b)
\end{tabular}
\end{minipage}
\begin{minipage}[c]{0.5\textwidth}

\begin{tabular}{c}
\includegraphics[width=\linewidth]{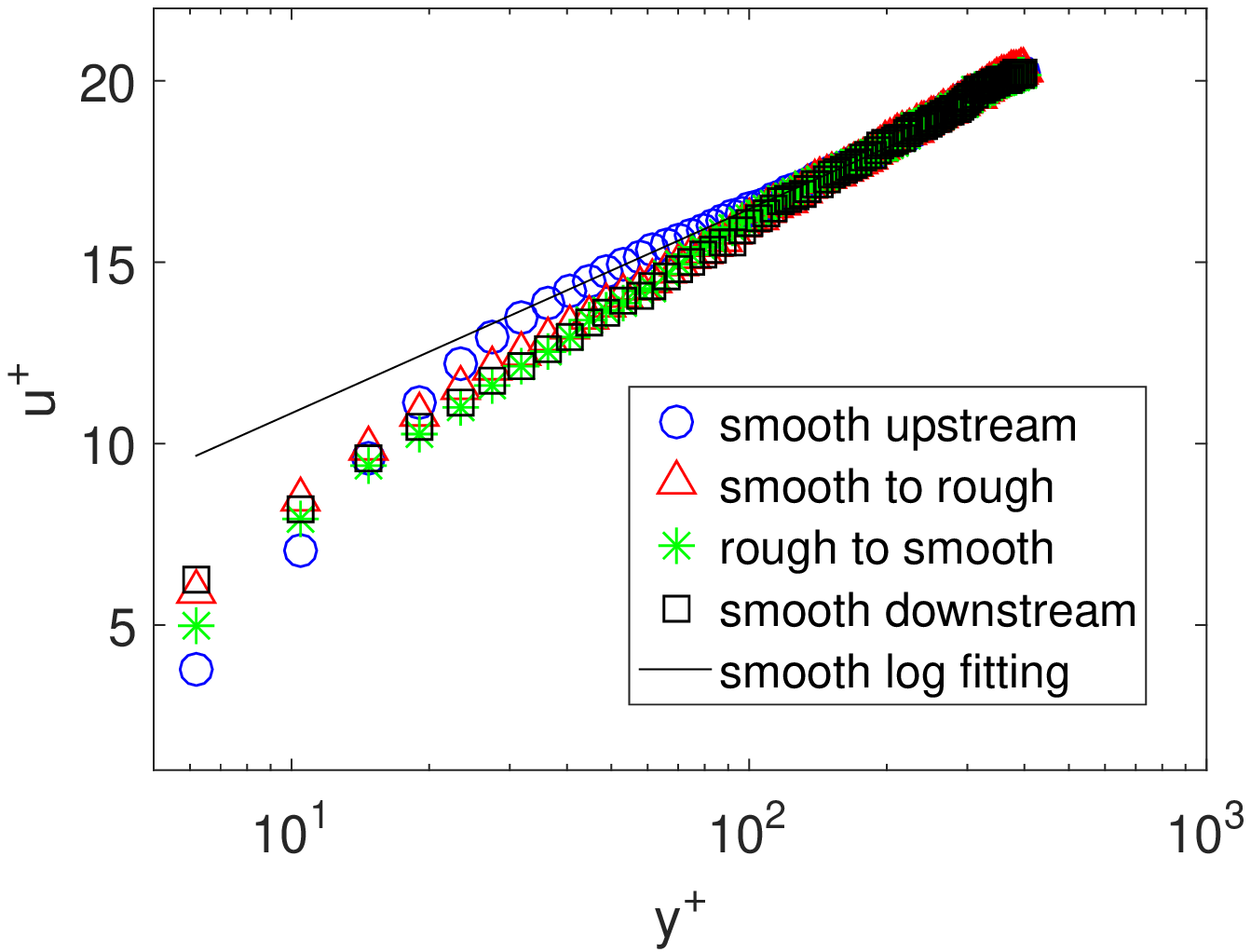}\\
      (c)
\end{tabular}
\end{minipage}
\begin{minipage}[c]{0.5\textwidth}
\hspace{\fill}
\begin{tabular}{c}
\includegraphics[width=\linewidth]{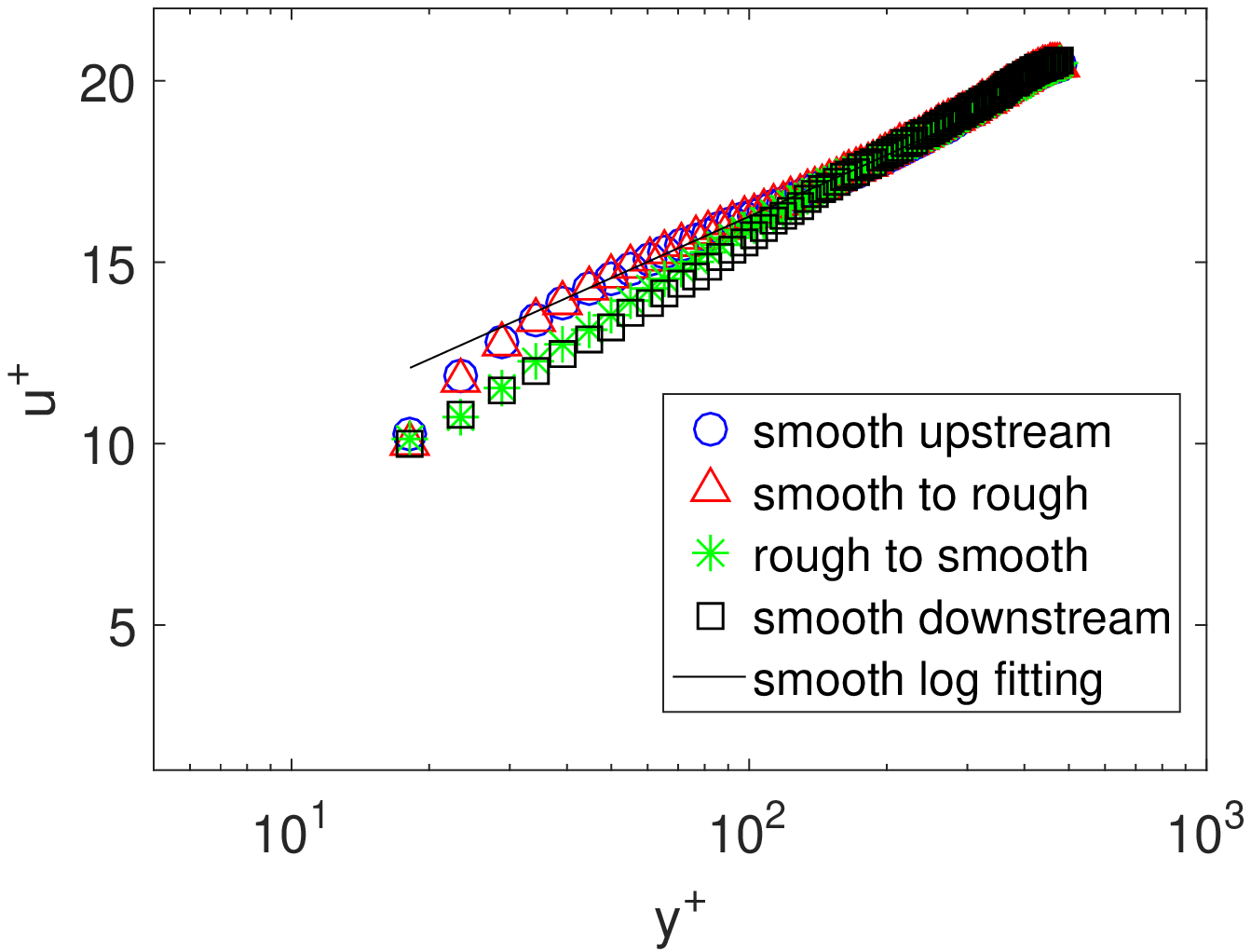}\\
      (d)
\end{tabular}
\end{minipage}
\caption{Mean velocity profiles in log-normal scales using the $y$ coordinate, for both flow rates and roughness types. The circles, triangles, asterisks and squares correspond to the flow over the smooth wall upstream of the rough elements, over the transition from the smooth wall to the rough surface, over the transition from the rough surface to the smooth wall, and over the smooth wall downstream of the rough elements, respectively. The continuous line corresponds to a fit for the overlap layer of the smooth wall profiles. (a) $Re = 7.8\cdot 10^{3}$ and d-type; (b) $Re = 9.6\cdot 10^{3}$ and d-type; (c)  $Re = 7.8\cdot 10^{3}$ and k-type; (d) $Re = 9.6\cdot 10^{3}$ and d-type.}
\label{fig:semilog_y}
\end{figure}

Figure \ref{fig:semilog_y} shows the mean velocity profiles in the traditional log-normal scales over the transition from the smooth wall to the rough surface and back to the smooth wall, for both flow rates. The origin of the vertical coordinate $y$ is the vertical position of the smooth surface. Figures \ref{fig:semilog_y}a and \ref{fig:semilog_y}b correspond to d-type roughness and Figs. \ref{fig:semilog_y}c and \ref{fig:semilog_y}d to k-type roughness, and the symbols used in the figures are explained in the key.

With the origin of $y$ at the top surface of rough elements, the mean velocities over the d-type rough elements increase or remain the same in the lower part of the profiles when compared to the flow over the upstream smooth wall, and then decrease over the smooth wall downstream of the rough elements to approximately the same values existing upstream of the elements. The higher velocities found over the d-type rough elements are due to stable inner-cavities vortices, which generate a positive velocity at $y=0$ over the cavities. The situation is different for the k-type roughness, for which the mean velocities remain the same or decrease over the rough elements when compared to the flow over the upstream and downstream smooth walls. Far from the walls, in the $y^+ > 100$ region, the mean velocity profiles collapse: the perturbations caused by the rough elements do not reach the upper regions of the flow within the longitudinal length of the rough plate. For the present smooth-rough and rough-smooth transitions, the perturbations do not reach the external layer. 

\begin{figure}[ht!]
\begin{minipage}[c]{0.5\textwidth}
\begin{tabular}{c}
\includegraphics[width=\linewidth]{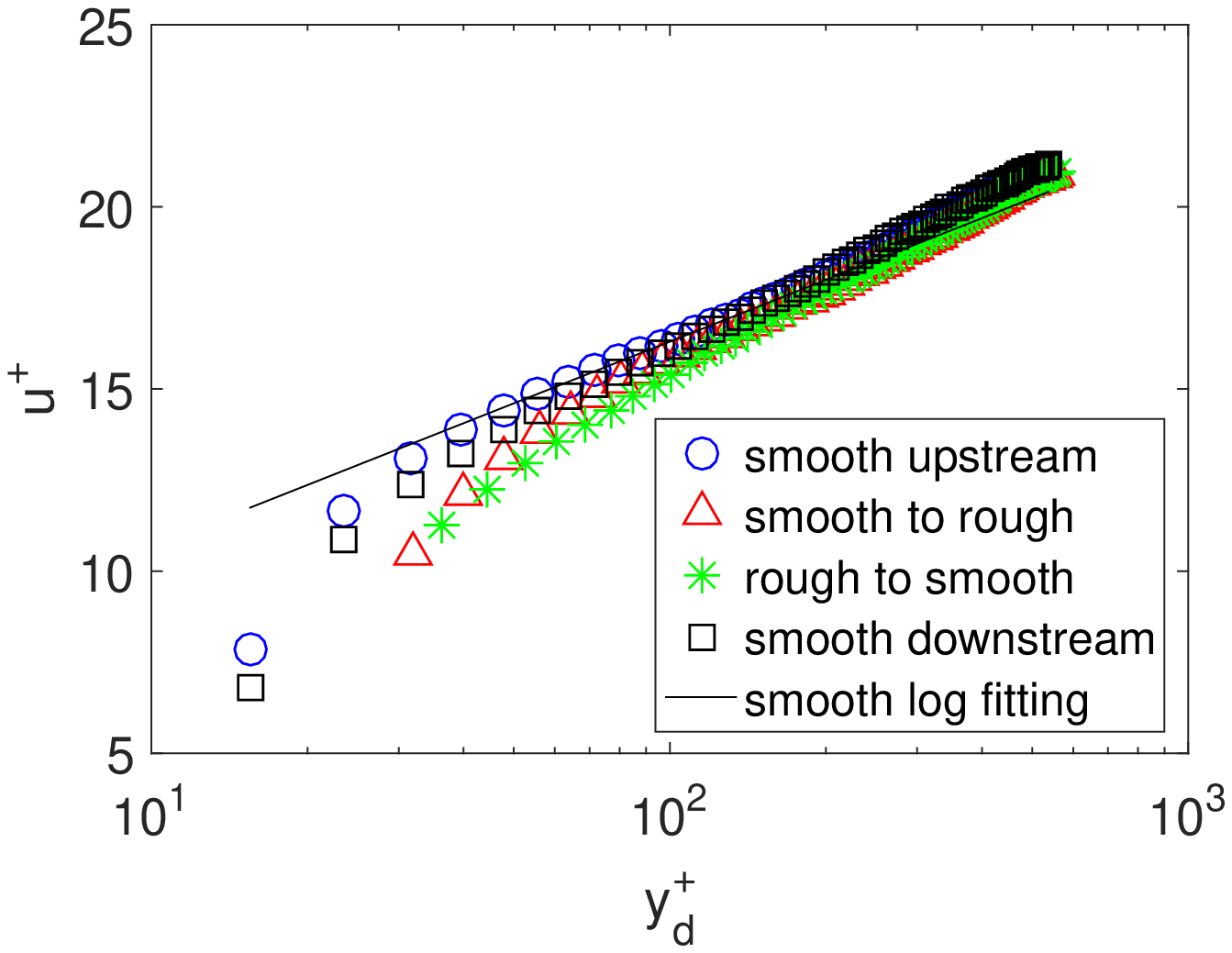}\\
      (a)
\end{tabular}
\end{minipage} \hfill
\begin{minipage}[c]{0.5\textwidth}
\hspace{\fill}
\begin{tabular}{c}
\includegraphics[width=\linewidth]{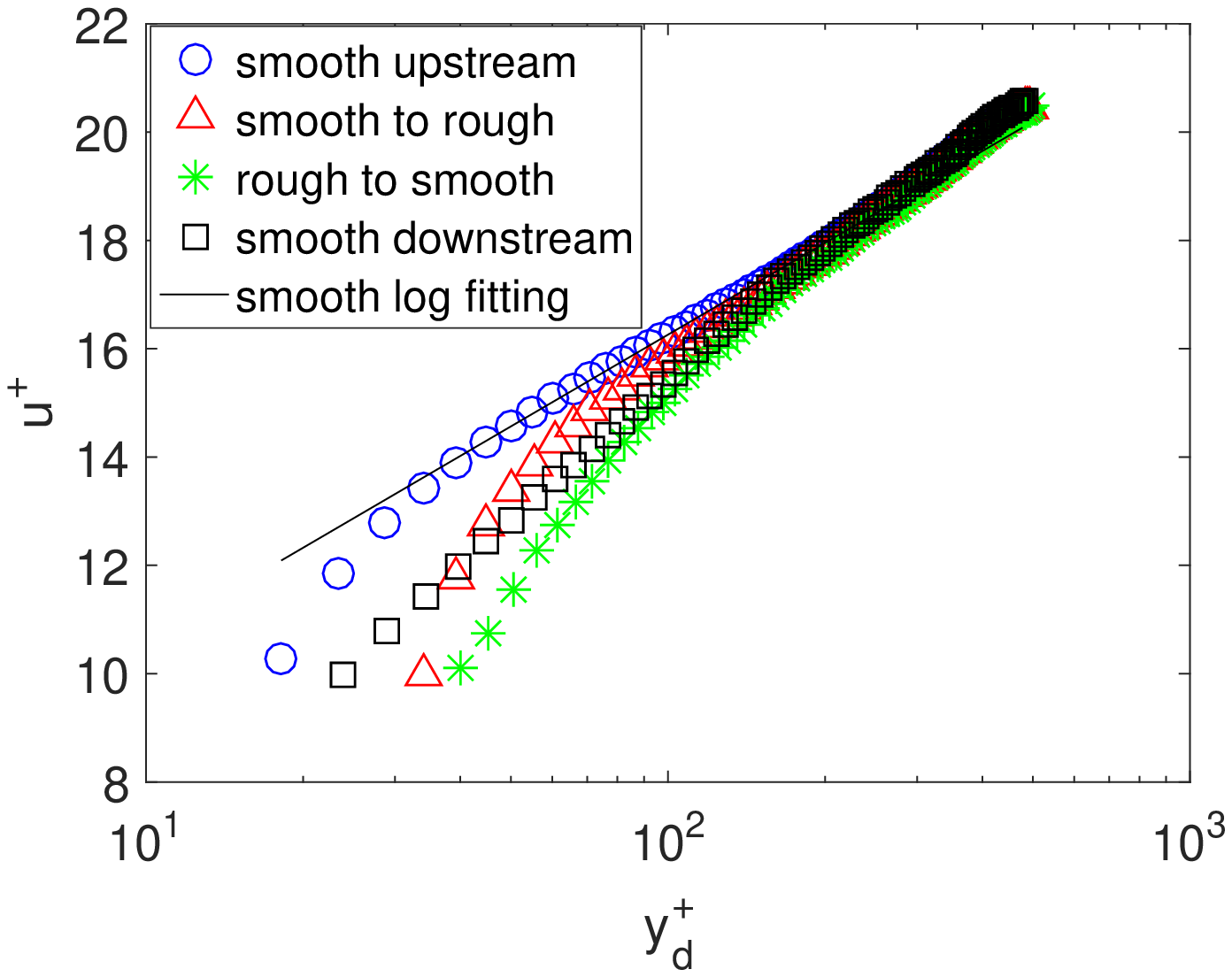}\\
      (b)
\end{tabular}
\end{minipage}
\begin{minipage}[c]{0.5\textwidth}

\begin{tabular}{c}
\includegraphics[width=\linewidth]{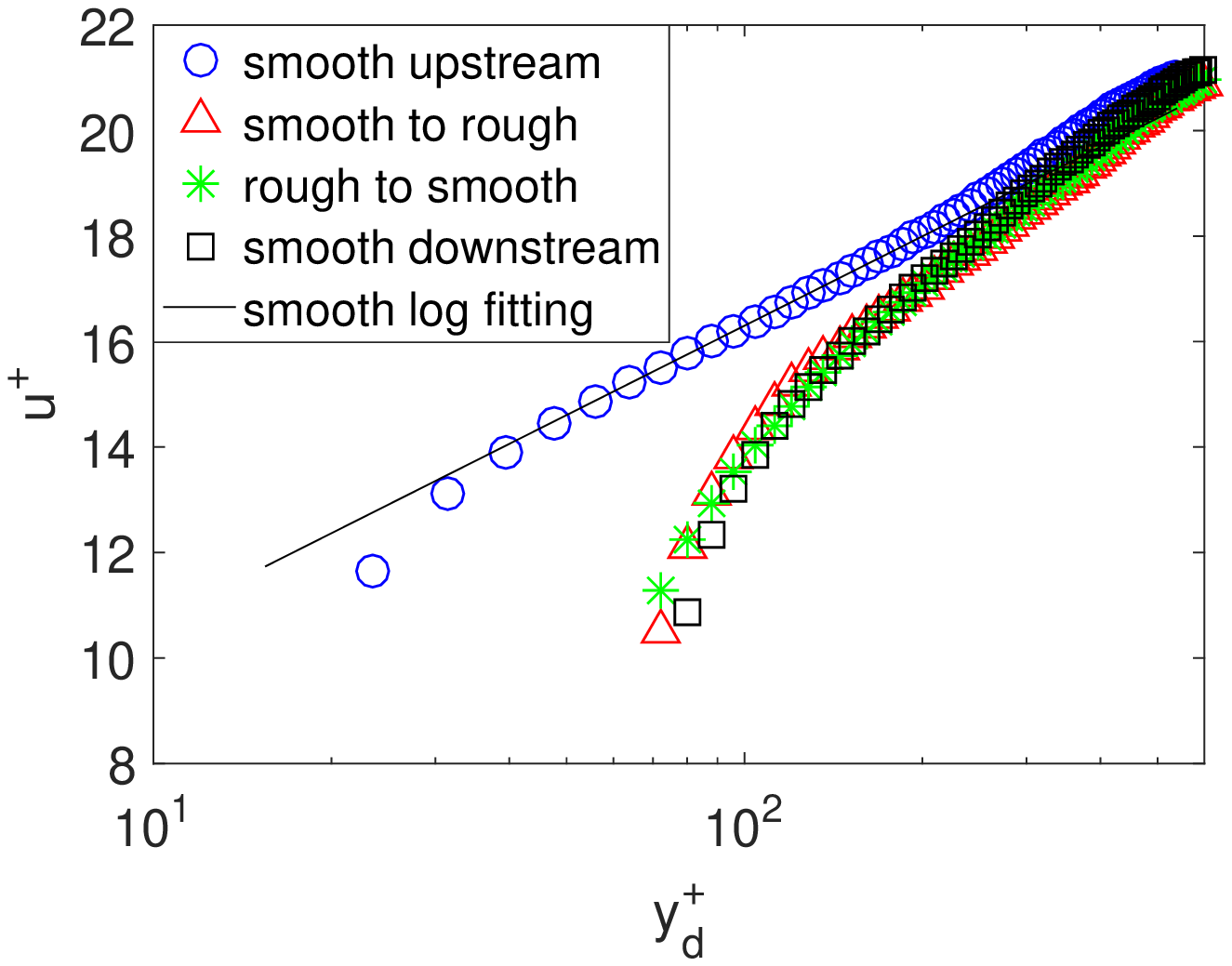}\\
      (c)
\end{tabular}
\end{minipage}
\begin{minipage}[c]{0.5\textwidth}
\hspace{\fill}
\begin{tabular}{c}
\includegraphics[width=\linewidth]{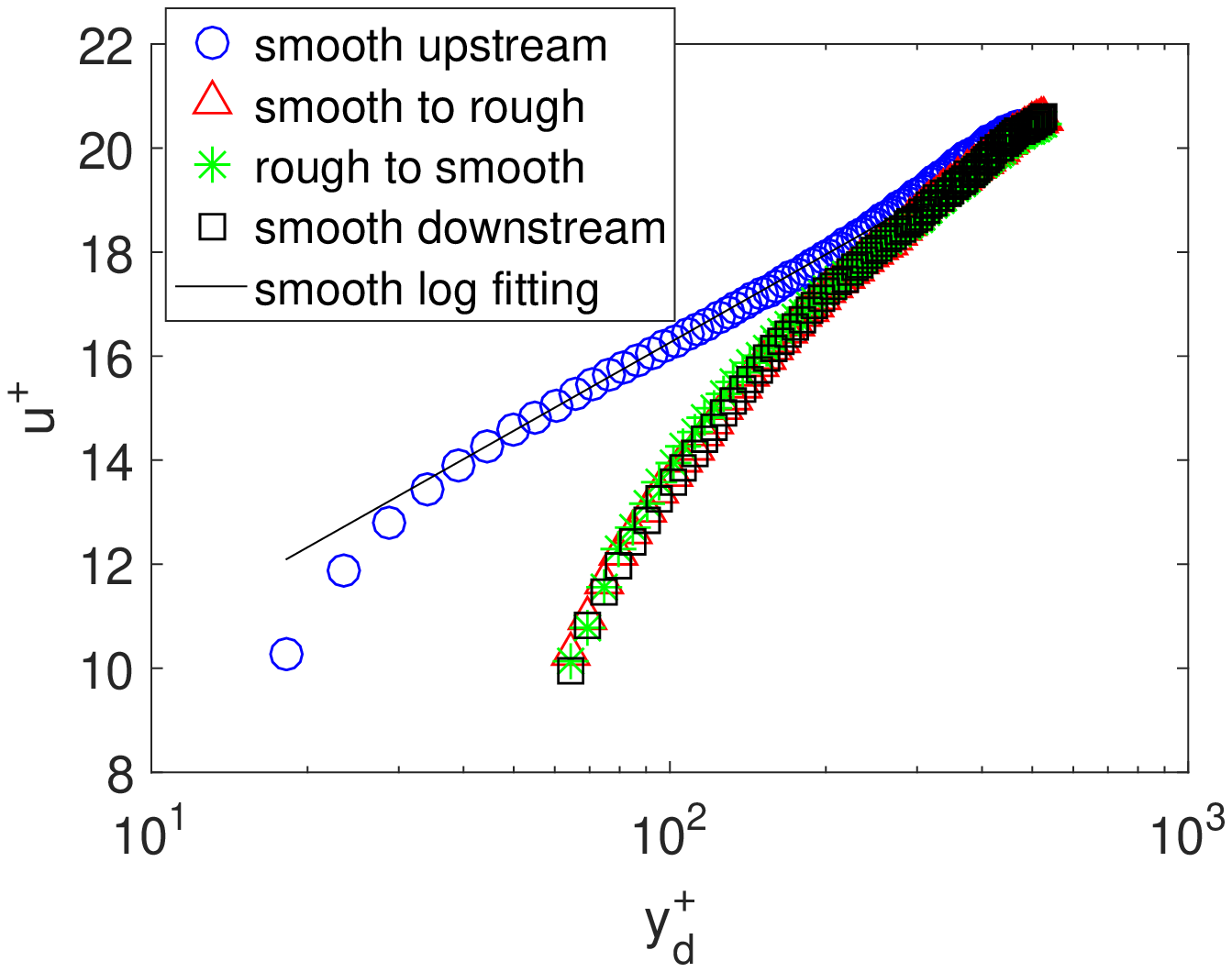}\\
      (d)
\end{tabular}
\end{minipage}
\caption{Mean velocity profiles in log-normal scales using the $y_d$ coordinate, for both roughness types and $Re = 9.6\cdot 10^{3}$. The circles, triangles, asterisks and squares correspond to the flow over the smooth wall upstream of the rough elements, over the transition from the smooth wall to the rough surface, over the transition from the rough surface to the smooth wall, and over the smooth wall downstream of the rough elements, respectively. The continuous line corresponds to a fit for the overlap layer of the smooth wall profiles. (a) d-type and $\overline{k}$; (b) k-type and $\overline{k}$; (c) d-type and $y_0$; (d) k-type and $y_0$.}
\label{fig:semilog_yd}
\end{figure}

The coordinate system with origin at the smooth wall surface, $y$, is commonly used in studies on boundary-layer transitions from smooth to rough walls \cite{Jimenez_1}. The use of this coordinate system close to the walls give the idea that higher velocities are obtained close to the walls; however, when the rough elements consist of cavities, the distance to the wall changes because wall material was removed. In order to analyze the evolution of mean velocities at the same distances from the walls, it is necessary to use an effective origin for the wall. For that, we use a displaced vertical coordinate, showed schematically in Fig. \ref{fig:geometry} and given by Eq. \ref{displaced_coordinate}:

\begin{equation}
	y_d = y  + \Delta y
	\label{displaced_coordinate}
\end{equation}

\noindent where $\Delta y$ is a distance below the smooth wall surface that corresponds to the effective position where the rough wall is. For fully-developed flows in rough regime, Nikuradse \cite{Nikuradse} showed that the effective height of rough walls is $y_0 = k/33$, so that $\Delta y = \left( k -y_0 \right)$ in Eq. \ref{displaced_coordinate}. If the flow is not fully developed, another possibility is to use the mean height weighted by the lengths and heights of elements, $\overline{k}$, and $\Delta y = \left( k -\overline{k} \right)$ in Eq. \ref{displaced_coordinate}.

Figure \ref{fig:semilog_yd} shows the mean velocity profiles using $y_d$ in the traditional log-normal scales over the transition from the smooth wall to the rough surface and back to the smooth wall, for $Re = 9.6\cdot 10^{3}$. Figures \ref{fig:semilog_y}a and \ref{fig:semilog_y}b correspond to d- and k-type roughness, respectively, with $\Delta y = \left( k -\overline{k} \right)$, and Figs. \ref{fig:semilog_y}c and \ref{fig:semilog_y}d to d- and k-type roughness, respectively, with $\Delta y = \left( k -y_0 \right)$. The symbols used in the figures are explained in the key. Using $y_d$, the mean velocities decrease over the rough elements when compared to the flow over the upstream and downstream smooth walls, and in the $y_d^+ > 100$ region for $\Delta y = \left( k -\overline{k} \right)$ and $y_d^+ > 200$ for $\Delta y = \left( k -y_0 \right)$ the mean velocity profiles collapse. Because the present case is a transition, the perturbations caused by the rough elements do not reach the upper regions of the flow. Therefore, in the $y_d$ coordinate, the presence of the rough elements always decrease the flow velocities close to the wall.

Although the effective height proposed by Nikuradse \cite{Nikuradse} is for fully-developed rough regimes, Figs. \ref{fig:semilog_y}c and \ref{fig:semilog_y}d show the smooth-rough transition from the perspective of a fully-rough regime. In a rough regime, the normalized log-normal profiles are parallel and displaced by a distance $\Delta B$ from the smooth profiles. In the present case, the flow is undergoing a transition from smooth to rough surfaces, and back to a smooth surface. Figs. \ref{fig:semilog_y}c and \ref{fig:semilog_y}d show that the mean flow is highly perturbed by the rough elements in the $y_d^+ < 100$ region and almost unperturbed in the $y_d^+ > 200$, meaning that the disturbances due to the rough elements could not reach the external layers of the flow.

\begin{figure}[ht]	
   \begin{minipage}[c]{.49\linewidth}
    \begin{center}
     \includegraphics[width=\linewidth]{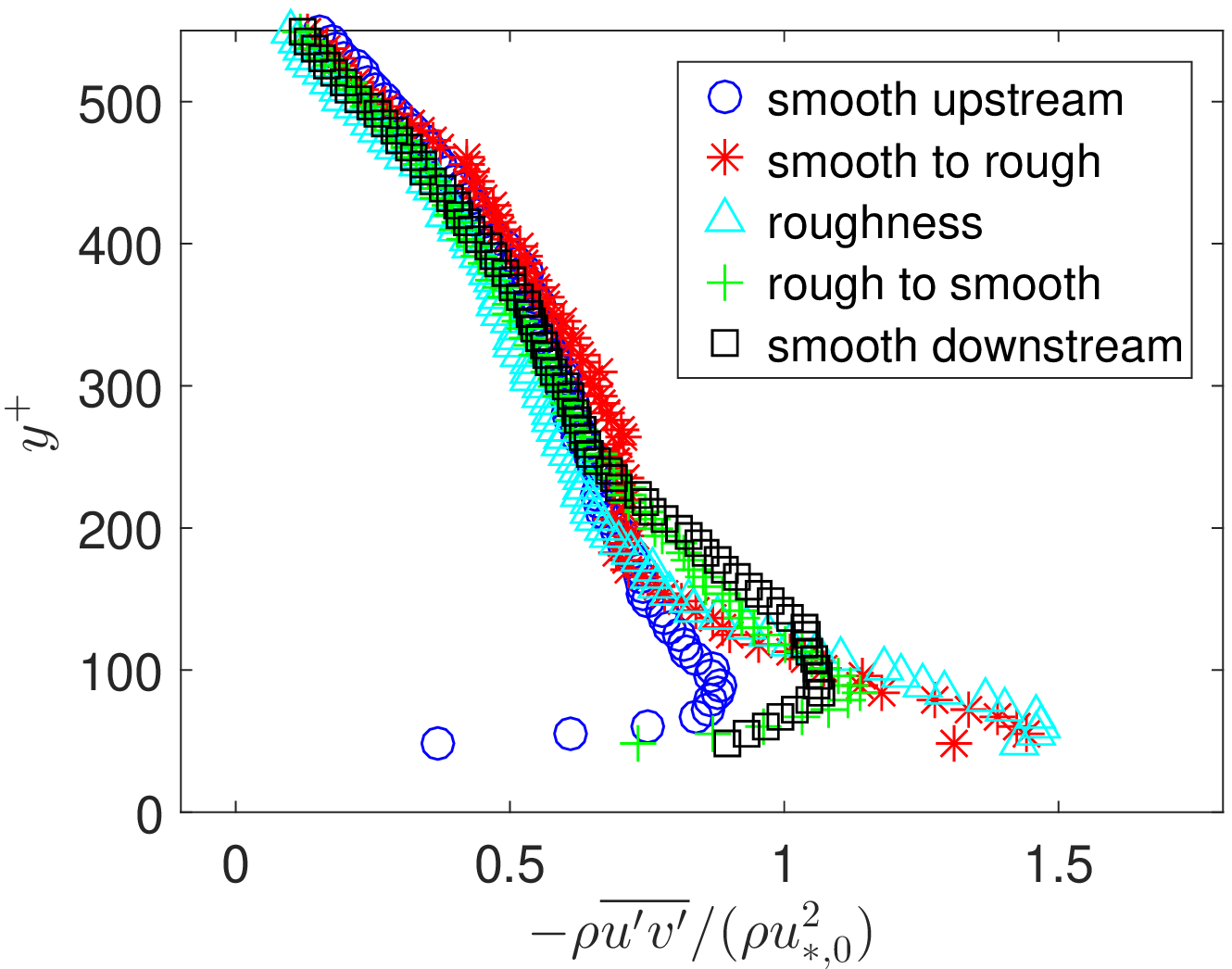}
		(a)
    \end{center}
   \end{minipage} \hfill
   \begin{minipage}[c]{.49\linewidth}
    \begin{center}
      \includegraphics[width=\linewidth]{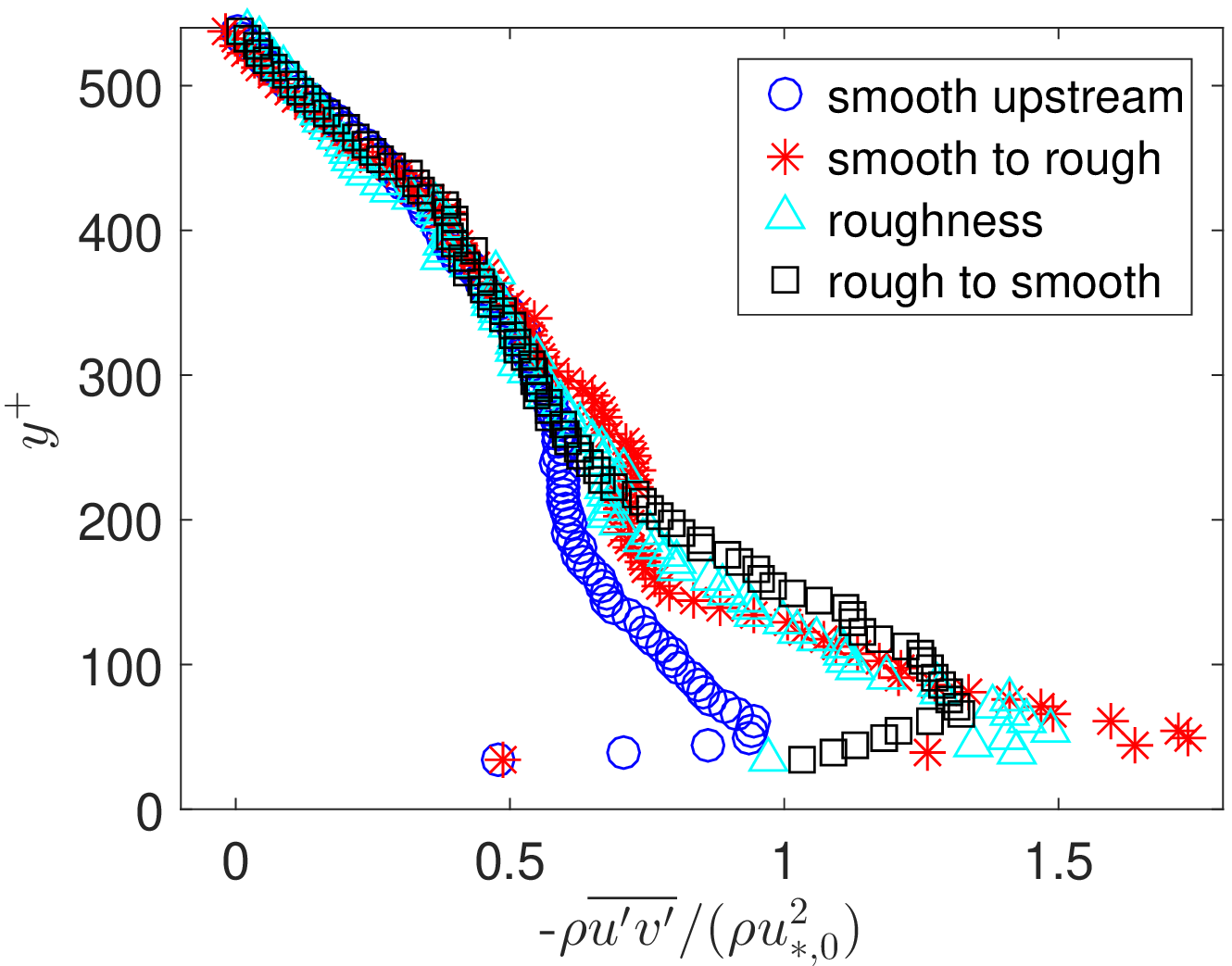}
			(b)
    \end{center}
   \end{minipage}
\caption{Mean profiles of $-\rho \overline{u'v'}$ normalized by $\rho u_{*,0}^2$ for $Re = 9.6\cdot 10^{3}$. (a) d-type roughness; (b) k-type roughness.}
\label{fig:turbulent_stress_rough}
\end{figure}

Figure \ref{fig:turbulent_stress_rough} shows the mean profiles of the $xy$ component of the turbulent stress normalized by $\rho u_{*,0}^2$ for $Re = 9.6\cdot 10^{3}$. Figure \ref{fig:turbulent_stress_rough}a corresponds to the d-type and Fig. \ref{fig:turbulent_stress_rough}b to the k-type roughness, respectively, and the symbols used in the figures are explained in the key. In order to make the profiles smoother, they were spatially averaged over a given longitudinal distance. For the d-type, the distances were of 5 rough elements in the transition from the smooth to the rough surfaces, 7 rough elements for the flow over the rough surface, and 3 rough elements for the transition from the rough to the smooth surfaces. For the k-type, the distances were of 3 rough elements in the transition from the smooth to the rough surfaces, 4 rough elements for the flow over the rough surface, and 1 rough element for the transition from the rough to the smooth surfaces.

In the longitudinal direction, $-\rho \overline{u'v'}$ increases near the bed as the flow goes from the smooth to the rough surfaces for both roughness types. The longitudinal increase in $-\rho \overline{u'v'}$ occurs in the $y^+ < 200$ region, and in the region $y^+ > 250$ the profiles collapse. In addition, the mean values of $-\rho \overline{u'v'}$ are higher over the smooth surface downstream of the rough elements than over the smooth surface upstream of the rough elements. The increase of $-\rho \overline{u'v'}$ over the rough region and just downstream of it is due to the interaction between vortices existing in the cavities and the flow in the upper regions. The existence of inner-cavities vortices is shown in Subsection \ref{subsection:temporal}. Because the flow is undergoing a smooth-rough transition in a limited length, the disturbances caused by the rough elements could not reach the external regions ($y^+ > 250$). 

\begin{figure}[ht]	
   \begin{minipage}[c]{.49\linewidth}
    \begin{center}
     \includegraphics[width=\linewidth]{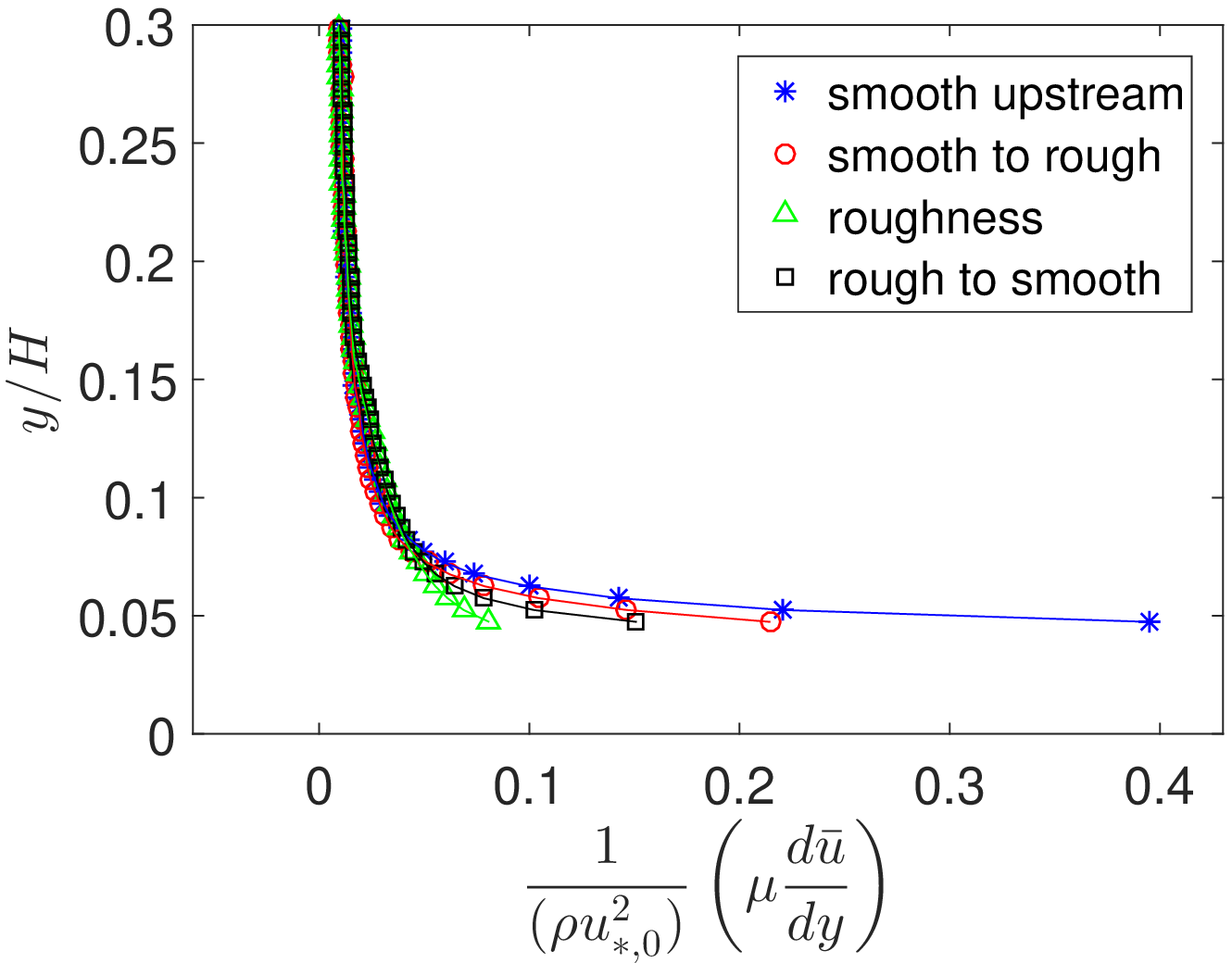}
		(a)
    \end{center}
   \end{minipage} \hfill
   \begin{minipage}[c]{.49\linewidth}
    \begin{center}
      \includegraphics[width=\linewidth]{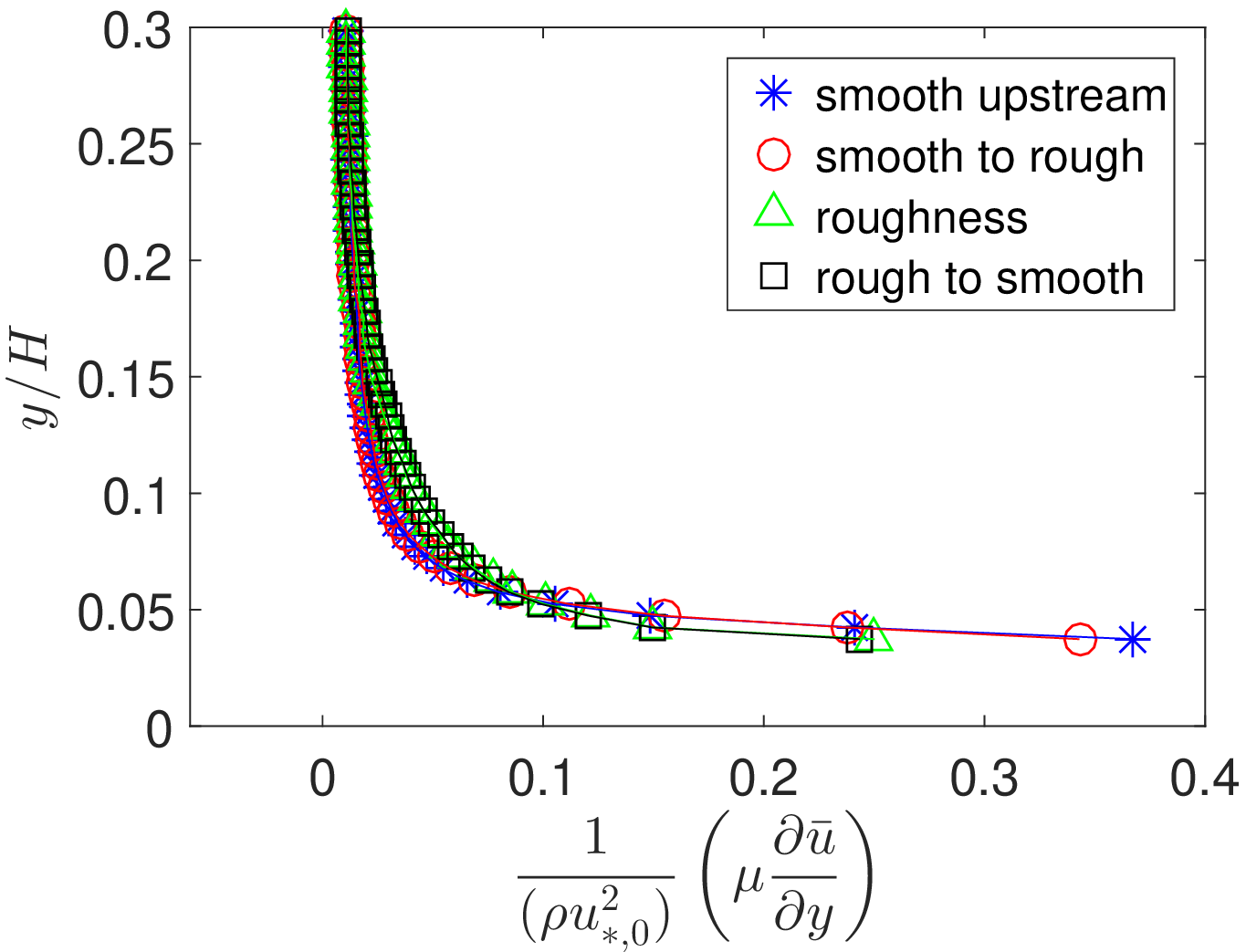}
			(b)
    \end{center}
   \end{minipage}
\caption{Mean profiles of the viscous stress normalized by $\rho u_{*,0}^2$ for $Re = 9.6\cdot 10^{3}$. (a) d-type roughness; (b) k-type roughness.}
\label{fig:viscous_stress_rough}
\end{figure}

Figure \ref{fig:viscous_stress_rough} shows the mean profiles of the viscous stress normalized by $\rho u_{*,0}^2$ for $Re = 9.6\cdot 10^{3}$. Figure \ref{fig:viscous_stress_rough}a corresponds to the d-type and Fig. \ref{fig:viscous_stress_rough}b to the k-type roughness, and the symbols used in the figures are explained in the key. The longitudinal positions of the profiles as well as the distances for the averages are the same as in Fig. \ref{fig:turbulent_stress_rough}. For both roughness types, $\mu d\overline{u}/dy$ is lower over the rough surfaces than over the smooth walls. The decrease in the viscous stress over the rough wall is due to the lower gradients of mean velocities that are caused by the presence of recirculation vortices in the cavities. The existence of inner-cavities vortices is shown in Subsection \ref{subsection:temporal}.

\begin{figure}[h!]
\begin{center}
	\begin{tabular}{c}
	\includegraphics[width=0.55\columnwidth]{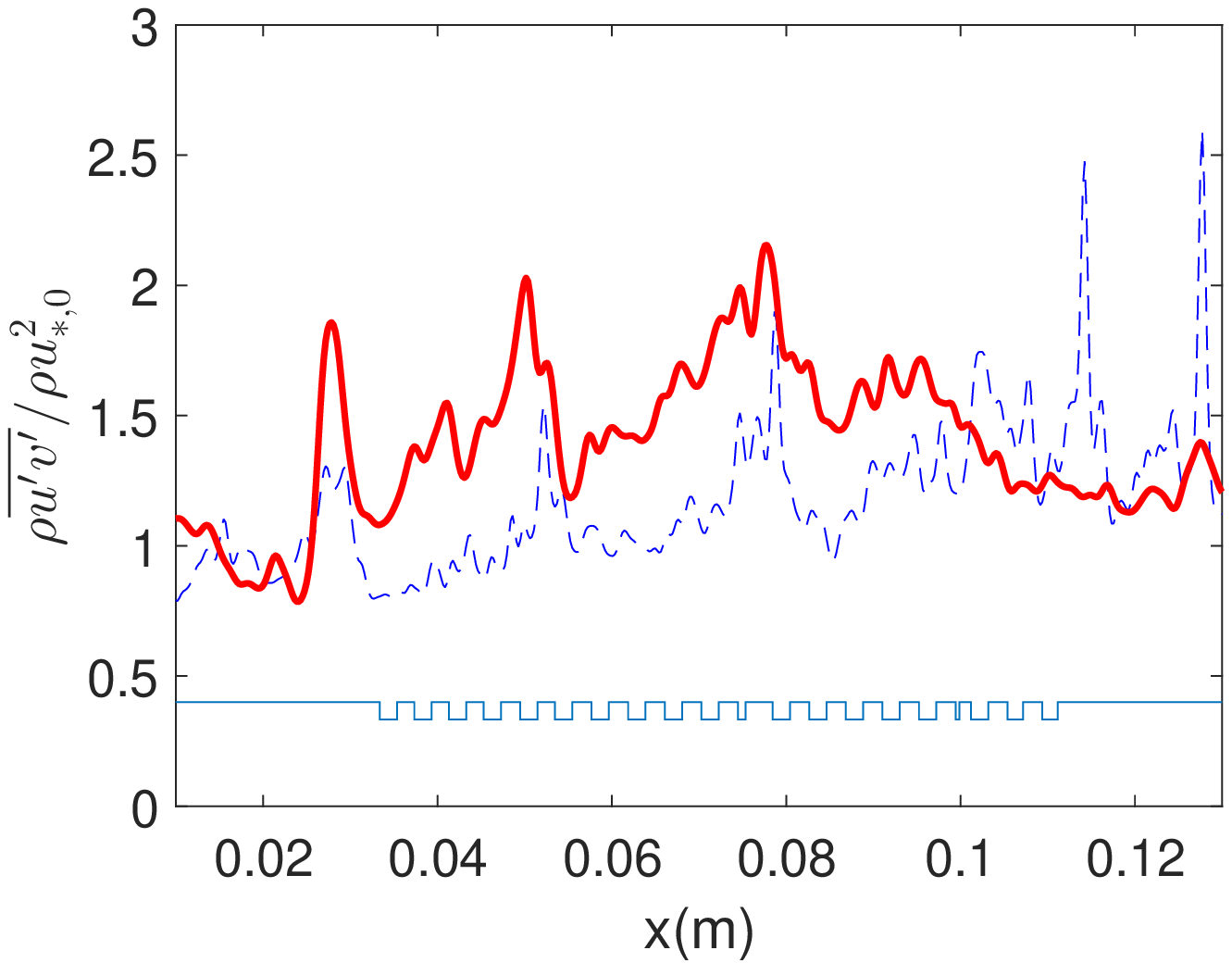}\\
	(a)\\
	\includegraphics[width=0.55\columnwidth]{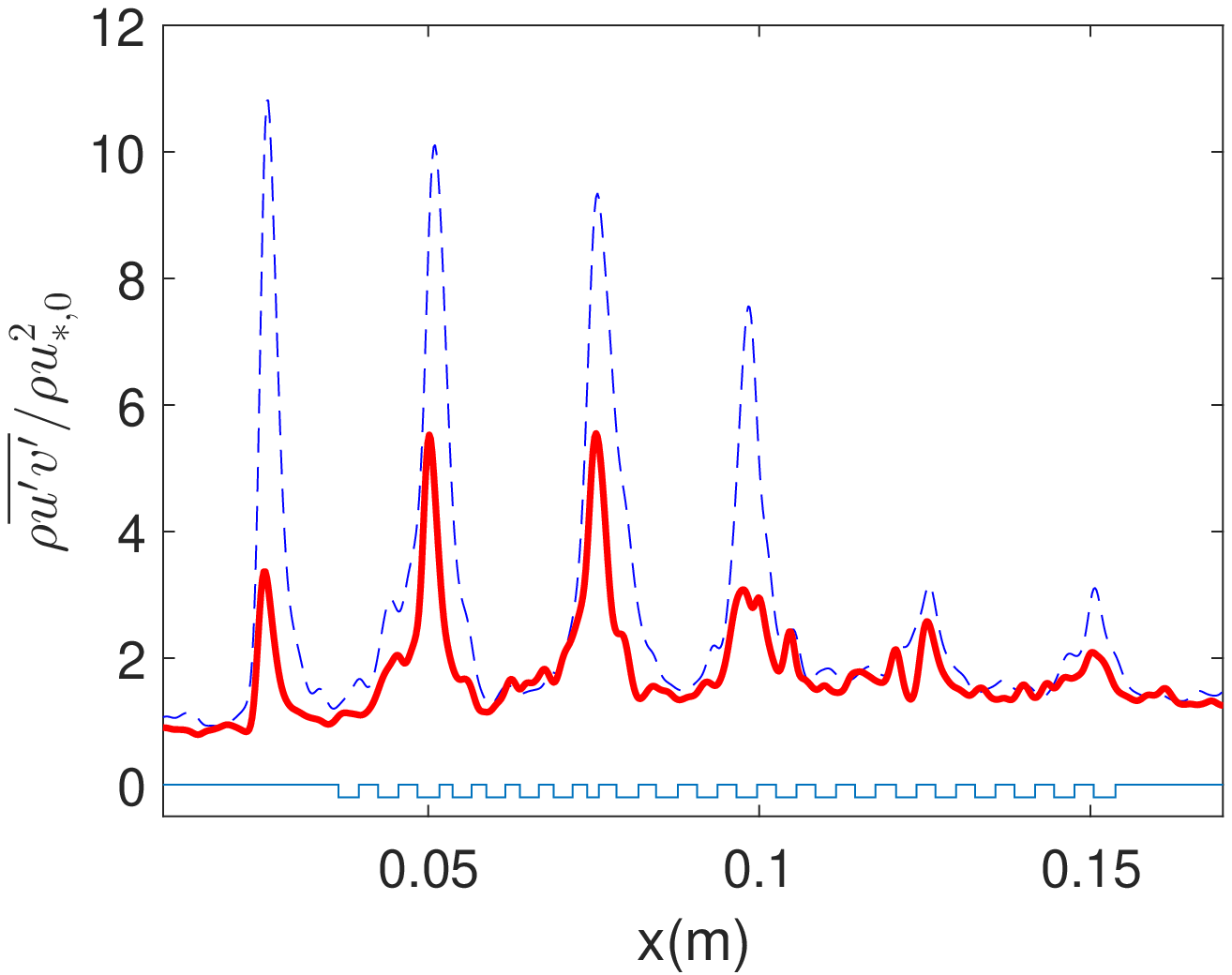}\\
	(b)
	\end{tabular} 
\end{center}
 	\caption{Maximum values of $-\rho \overline{u'v'}$ along the longitudinal direction normalized by $\rho u_{*,0}^2$. (a) d-type and (b) k-type rough elements. The dashed and continuous lines correspond to $Re = 7.8\cdot 10^{3}$ and $Re = 9.6\cdot 10^{3}$, respectively.}
 	\label{fig:max_RST}
\end{figure}

Figure \ref{fig:max_RST} shows the maximum values of each vertical profile of $-\rho \overline{u'v'}$ along the longitudinal direction normalized by $\rho u_{*,0}^2$. The dashed and continuous lines correspond to $Re = 7.8\cdot 10^{3}$ and $Re = 9.6\cdot 10^{3}$, respectively, and Fig. \ref{fig:max_RST}a corresponds to the d-type and Fig. \ref{fig:max_RST}b to the k-type roughness. The maximum values were obtained from the local profiles of $-\rho \overline{u'v'}$, and they were filtered with a Gaussian filter in order to eliminate noise.

In the transition from the smooth to the rough surfaces, and downstream of it, the maximum values of $-\rho \overline{u'v'}$ increase with respect to the smooth wall values. The increase of turbulent stresses over the rough surface is due to the formation of vortices in the cavities, and to the shedding of fluid from the cavities to the upper regions of the flow. In addition, Fig. \ref{fig:max_RST} shows the presence of oscillations in the maximum values of $-\rho \overline{u'v'}$, that are higher for the k-type roughness. With the exception of Lee \cite{Lee_B}, who presented a numerical study using DNS, these oscillation were never discussed in previous works and are experimentally shown here for the first time. In the case of Lee \cite{Lee_B}, the rough elements were sparse ($w/k = 8$) and with step change, and he found that there were overshoots in second order moments generated within the cavities or at the leading edges of rough elements. In the present case, the rough elements are more concentrated, so that the wavelength of oscillations is not the distance between rough elements. Instead, the oscillations have a wavelength $L$ within $5 w$ and $7 w$, and, as in the case of Lee \cite{Lee_B}, one possible explanation is the ejection of fluid from the cavities. The investigation of the inner-cavities vortices, presented in Subsection \ref{subsection:temporal}, corroborates the explanations based on the fluid ejection.

\begin{figure}[h!]
\begin{center}
	\begin{tabular}{c}
	\includegraphics[width=0.55\columnwidth]{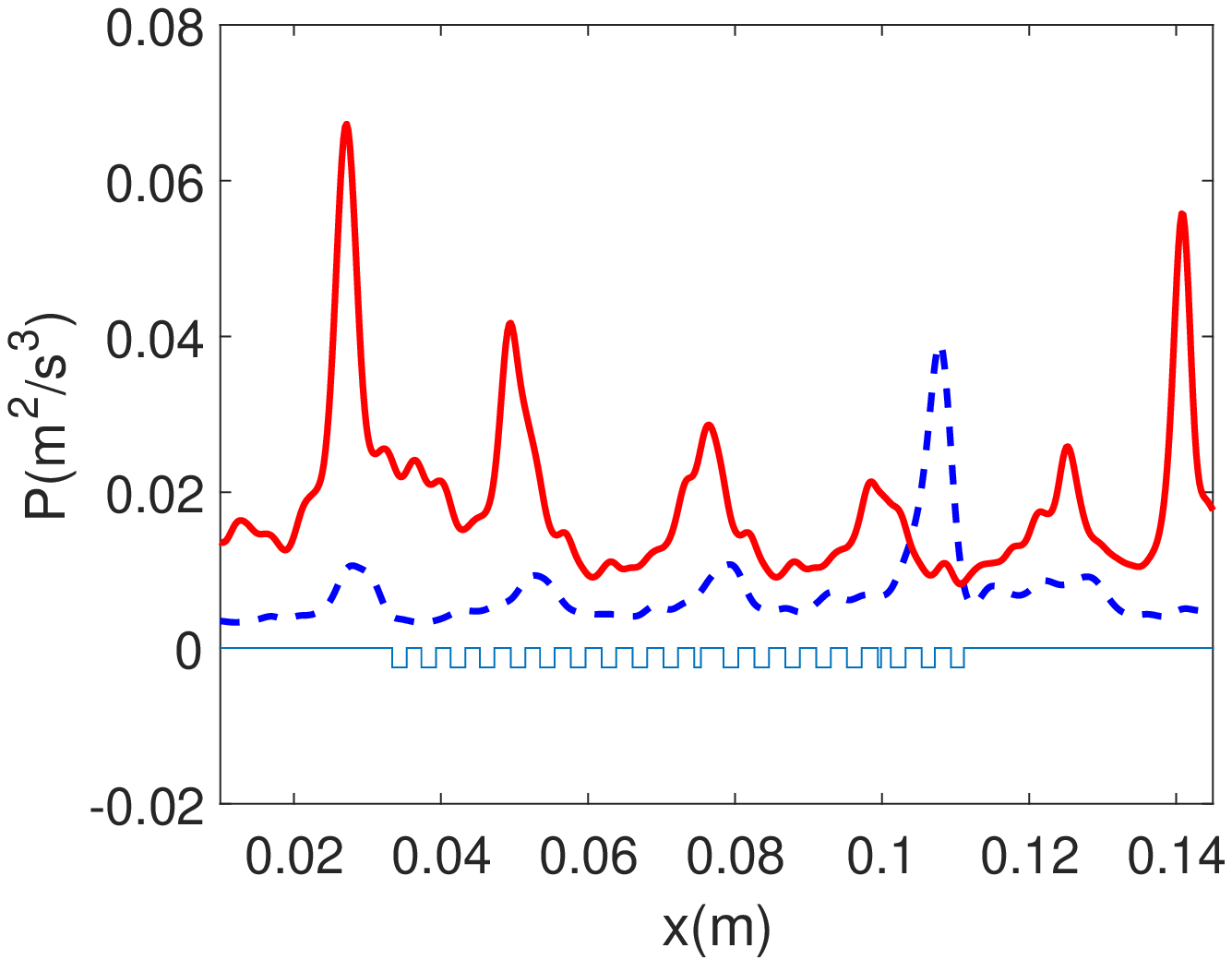}\\
	(a)\\
	\includegraphics[width=0.55\columnwidth]{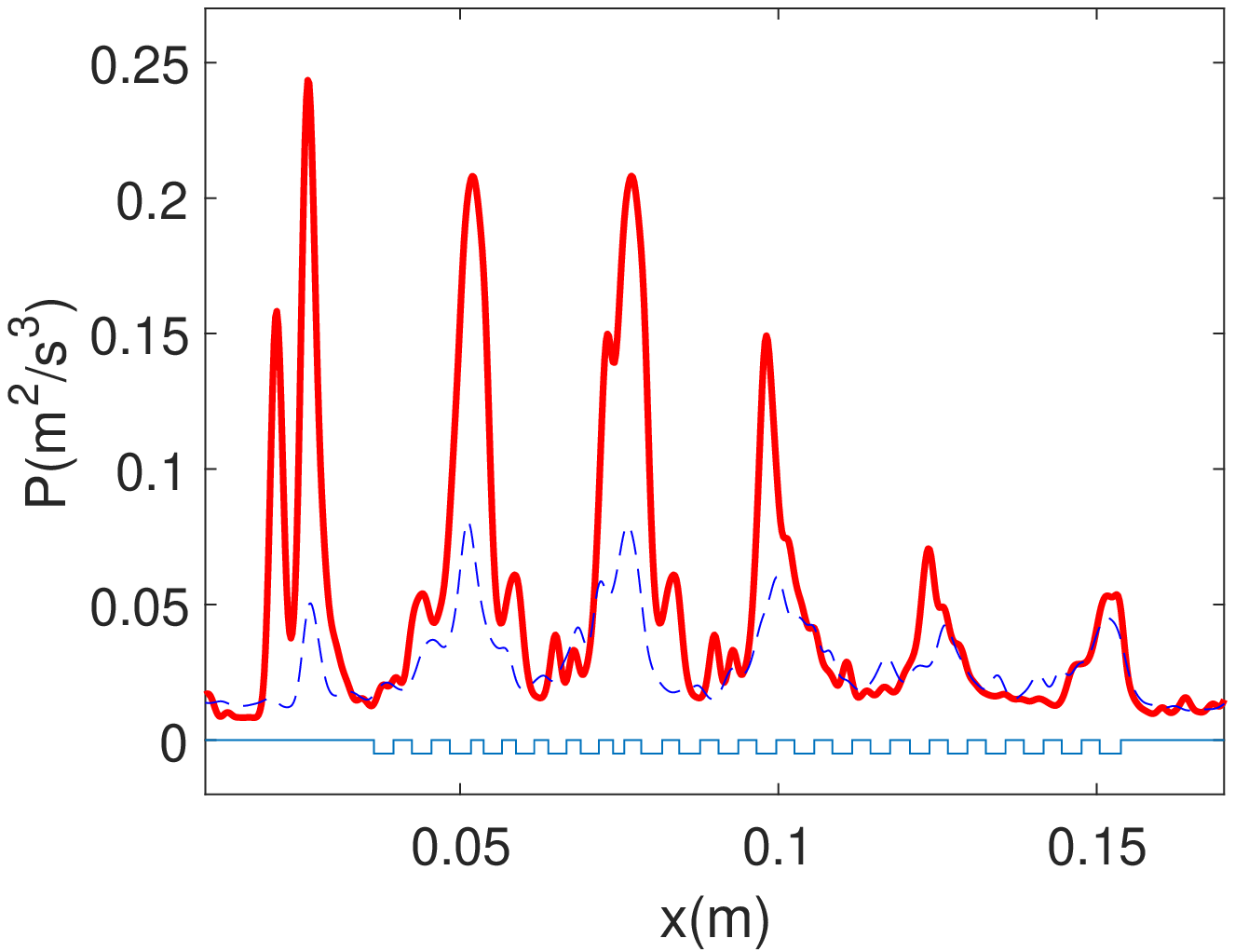}\\
	(b)
	\end{tabular} 
\end{center}
 	\caption{Maximum values of the $xy$ component of the turbulence production along the longitudinal direction. (a) d-type and (b) k-type rough elements. The dashed and continuous lines correspond to $Re = 7.8\cdot 10^{3}$ and $Re = 9.6\cdot 10^{3}$, respectively.}
 	\label{fig:max_prod_turb}
\end{figure}

Figure \ref{fig:max_prod_turb} shows the maximum values of the $xy$ component of the turbulence production along the longitudinal direction. The dashed and continuous lines correspond to $Re = 7.8\cdot 10^{3}$ and $Re = 9.6\cdot 10^{3}$, respectively, and Fig. \ref{fig:max_prod_turb}a corresponds to the d-type and Fig. \ref{fig:max_prod_turb}b to the k-type roughness. The $xy$ component of the turbulence production was computed as

\begin{equation}
	P = - \overline{u'v'} \frac{\partial \overline{u}}{\partial y}
	\label{prod_turb_eq}
\end{equation}

\noindent The maximum values were obtained from the local profiles of $P$, and they were filtered with a Gaussian filter in order to eliminate noise. As with the turbulent stress, the maximum values of $P$ increase with respect to the smooth wall values in the transition from the smooth to the rough surfaces, and downstream of it. Also, oscillations in $P$, that are higher for the k-type roughness, are present. The wavelength of the oscillations are the same as for $-\rho \overline{u'v'}$: $5w\, \leq\, L\, \leq\, 7w$. The similar behavior between $P$ and $-\rho \overline{u'v'}$ indicates that the Reynolds stresses are dominant in the turbulence production term in the region where the maxima occur. Because $-\rho \overline{u'v'}$ is the dominant term, we propose that the oscillations in turbulence production are also due to the ejection of fluid from the cavities.

\subsection{Rough walls: temporal analysis}
\label{subsection:temporal}

The flow in the cavities consists basically of vortices which are more or less stable depending on the roughness geometry. The dynamics of some inner-cavities vortices was investigated using high-speed flow visualization; therefore, the measurements are used for a temporal analysis. The movies showed the presence of vortices that intermittently eject mass with a given frequency to the upper regions, exchanging then mass with upper layers. From the high-speed movies, the angular velocities of inner-cavities vortices were computed with Matlab scripts written in the course of this work, and dominant frequencies are investigated.

\begin{figure}[h!]
\begin{center}
	\begin{tabular}{c}
	\includegraphics[width=0.55\columnwidth]{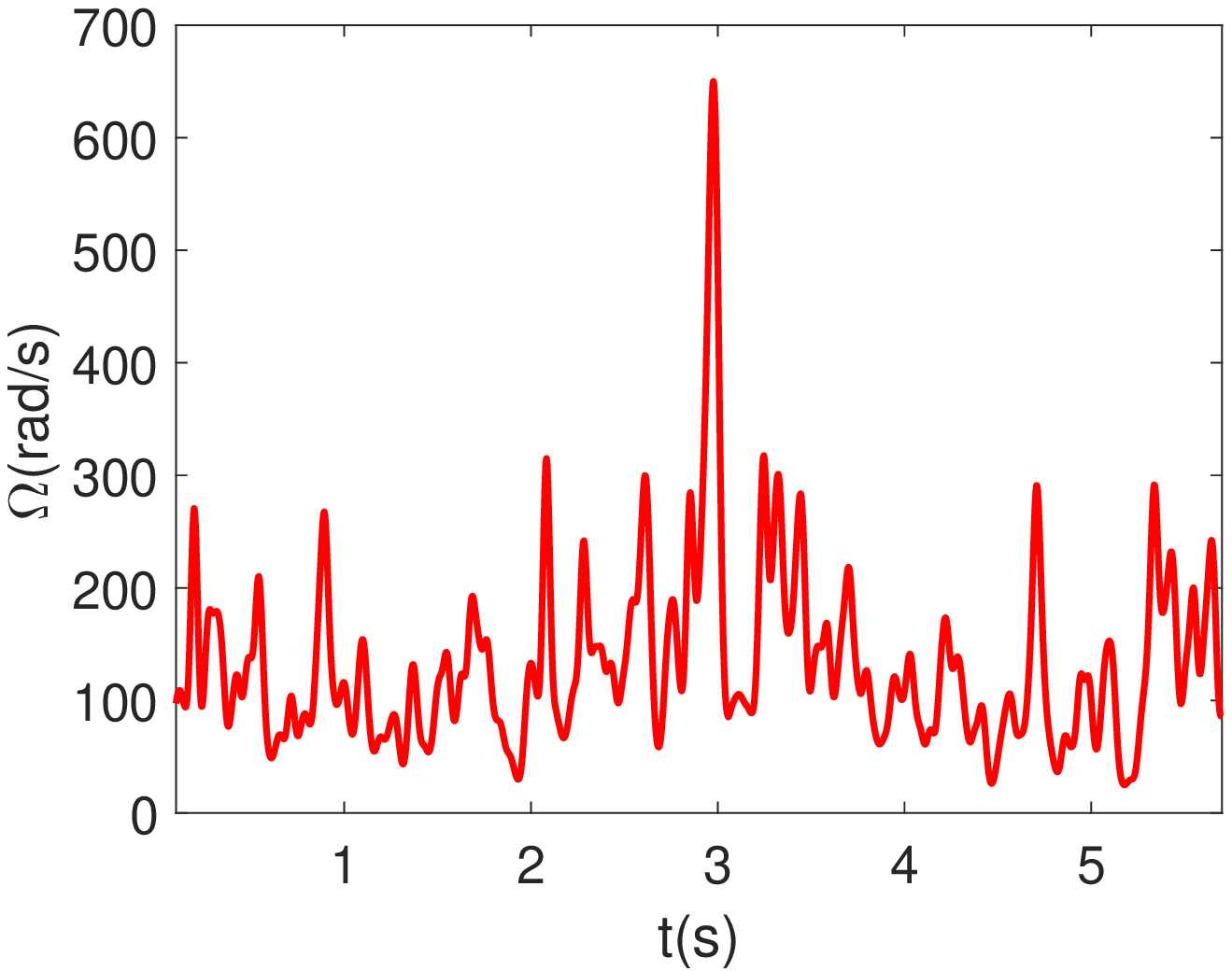}\\
	(a)\\
	\includegraphics[width=0.55\columnwidth]{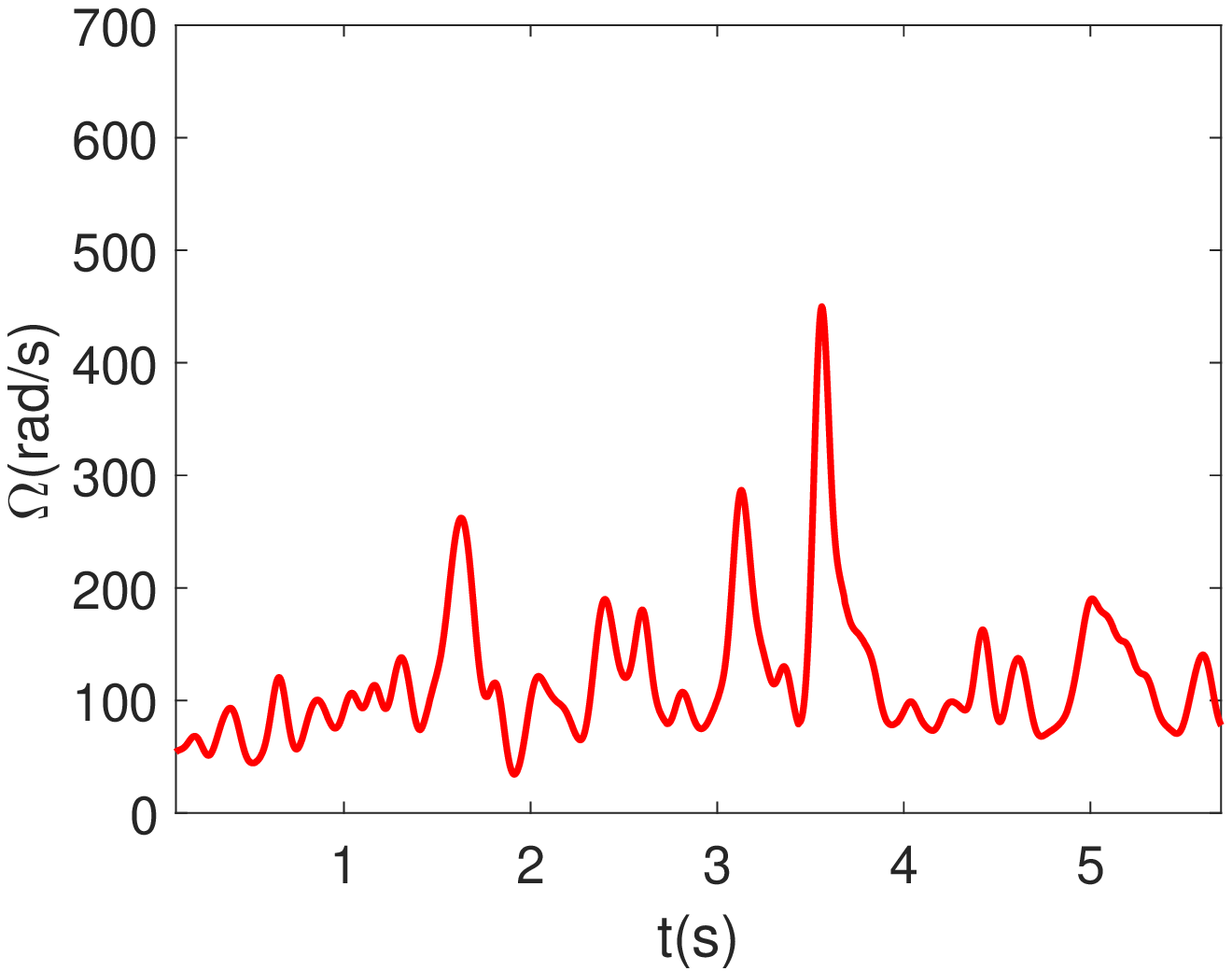}\\
	(b)
	\end{tabular} 
\end{center}
 	\caption{Angular velocities of inner-cavities vortices as function of time for $Re = 9.6\cdot 10^{3}$. (a) d-type; (b) k-type.}
 	\label{fig:angular velocity}
\end{figure}

Figure \ref{fig:angular velocity} shows the angular velocities of inner-cavities vortices along time for $Re = 9.6\cdot 10^{3}$. The data presented in Fig. \ref{fig:angular velocity} was filtered with a Gaussian filter in order to eliminate high frequency noise present in the measurements. Figure \ref{fig:angular velocity}a corresponds to the d-type and Fig. \ref{fig:angular velocity}b to the k-type roughness. From the present data, the mean angular velocities are $\overline{\Omega} \approx 130$ rad/s and $\overline{\Omega} \approx 120$ rad/s for the d- and k-types, respectively, which correspond to mean tangential velocities $\overline{v}_t = \overline{\Omega} / r$ of 0.06 and 0.08 m/s, respectively, where $r$ is the distance from the tracked particle to the center of the vortex. To investigate the dominant frequencies in the inner-cavities vortices, a power spectrum analysis of the angular velocities was made based on Hilbert transforms.

\begin{figure}[h!]
\begin{center}
	\begin{tabular}{c}
	\includegraphics[width=0.55\columnwidth]{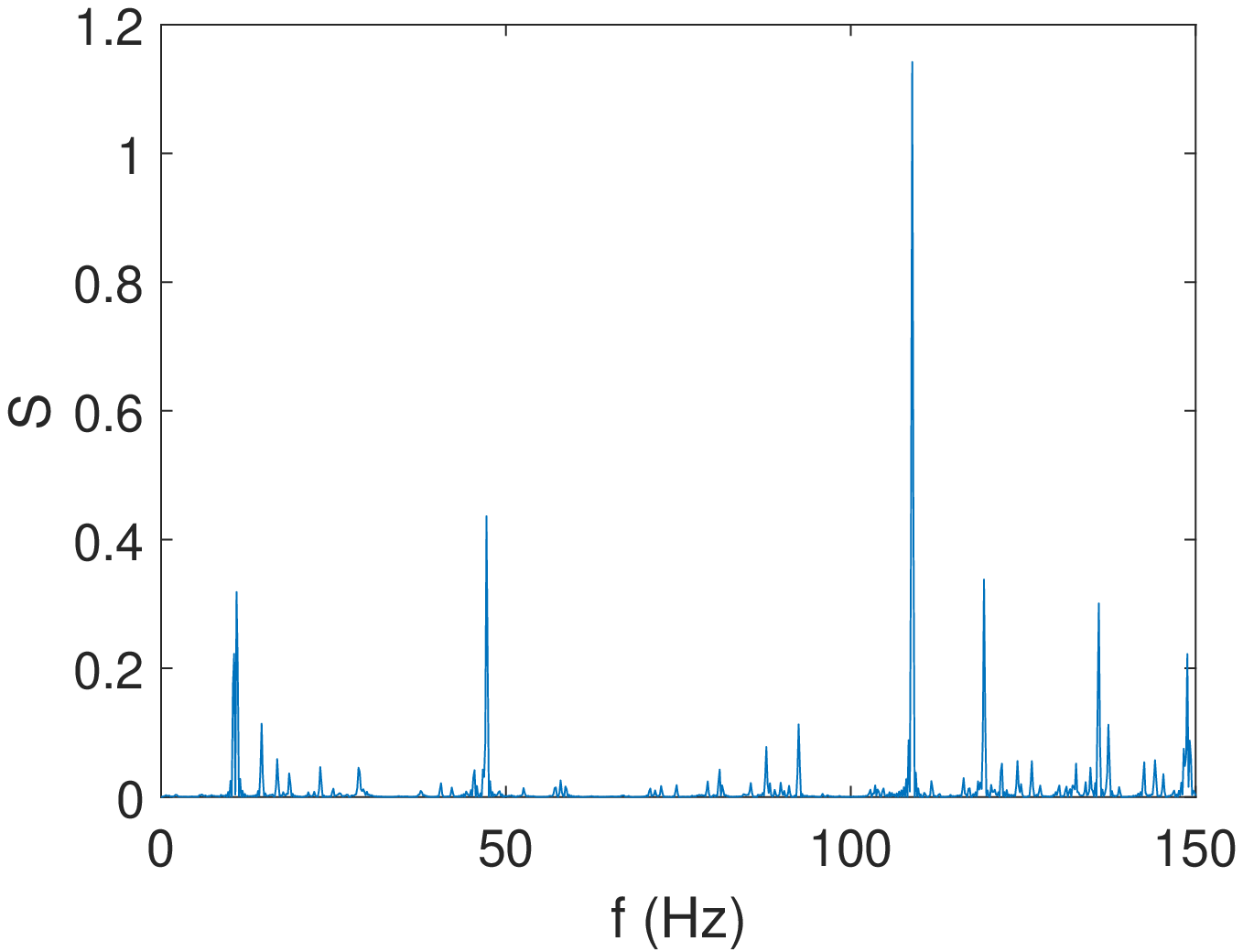}\\
	(a)\\
	\includegraphics[width=0.55\columnwidth]{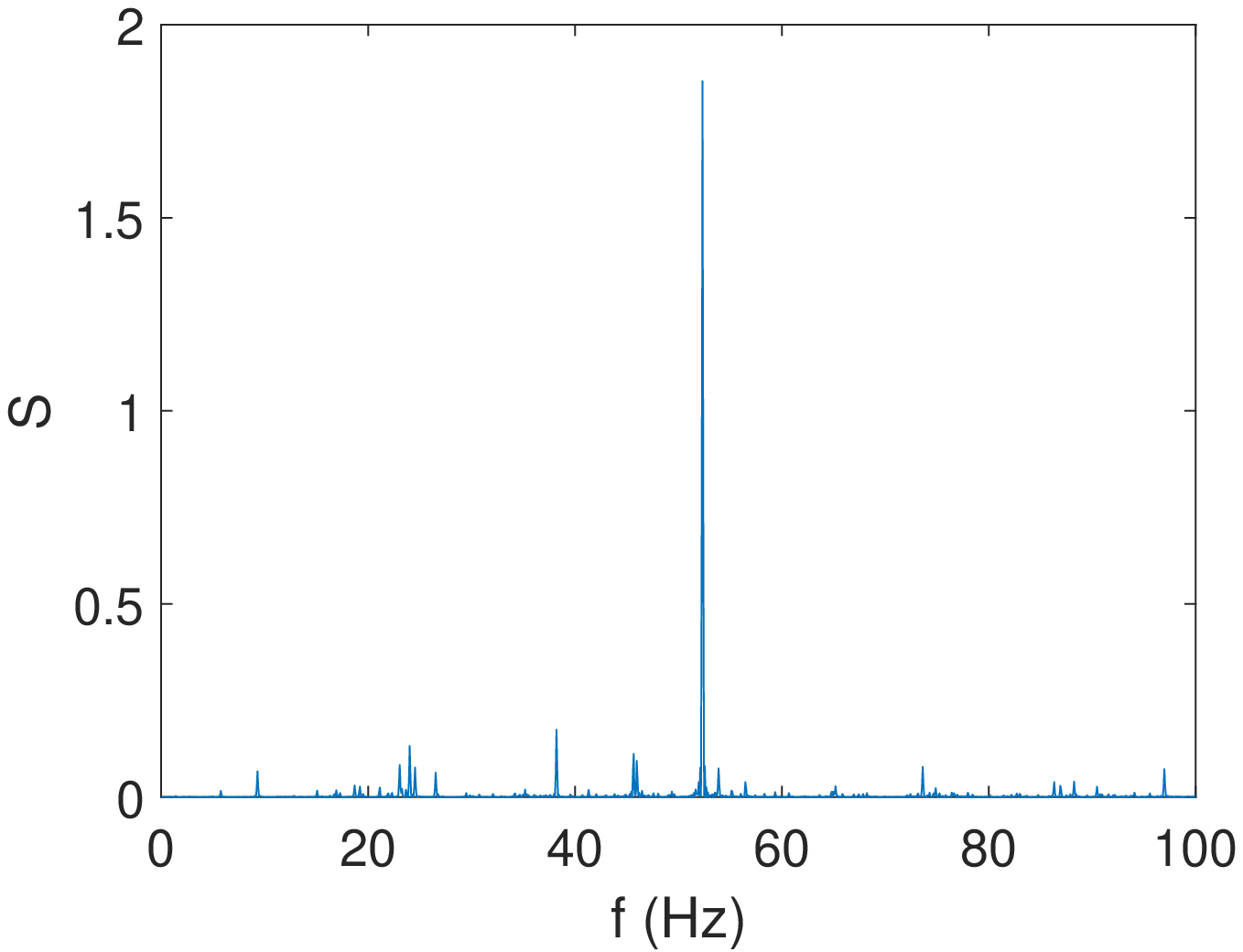}\\
	(b)
	\end{tabular} 
\end{center}
 	\caption{Power spectrum of the angular velocities based on the non-filtered data, for $Re = 9.6\cdot 10^{3}$. (a) d-type; (b) k-type.}
 	\label{fig:power_spectrum}
\end{figure}

Figure \ref{fig:power_spectrum} presents the power spectrum of the angular velocities based on the non-filtered data, for $Re = 9.6\cdot 10^{3}$. Figure \ref{fig:power_spectrum}a corresponds to the d-type and Fig. \ref{fig:power_spectrum}b to the k-type roughness. Figure \ref{fig:power_spectrum} shows the existence of a dominant frequency $f$ for the inner-cavities vortices, around 100 Hz for the d-type and 50 Hz for the k-type roughness, that are stronger for the k-type roughness. Based on $\overline{v}_t$, the vortices wavelengths are computed as $L = \overline{v}_t / f$, corresponding to wavelengths of approximately 0.3 $w$ and 0.5 $w$ for the d- and k-types, respectively. We note that those wavelengths are one tenth of the wavelengths of longitudinal oscillations of turbulent stress and turbulence production observed in the upper layers.

Although a theoretical explanation for the factor 10 is not advanced in this paper, we believe that there is a correlation between the angular velocities of the inner-cavities vortices and the flow oscillations in the layers above the rough elements. From the movies, we observed that fluid ejections from the cavities to the upper layers of the flow occur at approximately every 10 turns of the vortices. However, the number of turns cannot be precisely accessed from the particle tracking velocimetry used in the visualization experiments, and the value 10 is a rough estimation from the movie frames. Considering the mean tangential velocities $\overline{v}_t$, the frequencies of ejection correspond to wavelengths of approximately 3 $w$ and 5 $w$ for the d- and k-type roughness, respectively, which are of the same order of the wavelengths observed in the upper layers.

\section{Conclusions}

This paper presented an experimental study on the transition from smooth to rough walls, and back to the smooth one, in turbulent closed-conduit flows. Smooth plates were inserted in the entrance length, a rough plate with smooth and rough sections was inserted in the test section, and a smooth plat was inserted in the final length of a closed conduit of rectangular cross section. Each rough section consisted of 20 square rough elements aligned in a direction transverse to the water flow, and were of d- or k-types. The experiments were performed at moderate Reynolds numbers ($Re$ = ord$(10^4)$), a case for which experimental data are scarce and the physics is not completely understood. The closed-conduit flow just upstream of the roughness transition was fully developed and hydraulically smooth, and the flow field was measured by low frequency PIV (particle image velocimetry) and by high-speed visualization.

The decorrelated instantaneous flow fields in the region above the rough elements were measured with PIV, from which the mean velocities and fluctuations were determined for both roughness types and flow rates. From these, the $xy$ components of viscous stress, Reynolds stress and turbulence production were computed, and the flows over different smooth and rough regions were compared.

Considering the origin of the vertical coordinate at the top surface of rough elements, we showed that mean velocities over the d-type rough elements increase in the lower part of the profiles when compared to the flow over the upstream smooth wall, while over k-type roughness they decrease. The higher velocities found over the d-type rough elements are due to stable inner-cavities vortices, that induce a non-zero velocity at the origin. Considering the origin of the vertical coordinate as proposed by Nikuradse \cite{Nikuradse} for fully-developed rough regimes, the mean velocities decrease over the rough elements of both types when compared to the flow over the upstream and downstream smooth walls. For both coordinate systems, the perturbations caused by the rough elements do not reach the upper regions of the flow within the longitudinal length of the rough plate.

In the transition from the smooth to the rough surfaces, and downstream from it, the maximum values of the $xy$ components of turbulent stress and turbulence production increase with respect to the upstream smooth wall values. We propose that this increase is due to the presence of vortices in the cavities, that shed fluid from the cavities to the upper regions of the flow. The results showed the presence of longitudinal oscillations in Reynolds stress and turbulence production, that are higher for the k-type roughness. The existence of longitudinal oscillations was experimentally shown for the first time in this paper. Their wavelength is between $5 w$ and $7 w$ for both flow rates and roughness types, and we propose that they are related to the ejection of fluid from the cavities.

A flow visualization device was conceived in the course of this work and consisted basically of a continuous 0.1 W laser, a high-speed camera, and Matlab scripts written by the authors. From the high-speed movies, fluid ejection from the cavities to upper regions was observed, and the angular velocities of inner-cavities vortices was determined by image processing. The spectral analysis of the angular velocities showed a dominant frequency for the inner-cavities vortices, that are stronger for the k-type roughness. The dominant frequency is around 100 Hz for the d-type and around 50 Hz for the k-type roughness, corresponding to wavelengths of 0.3 $w$ and 0.5 $w$, respectively, which are one tenth of the wavelengths of turbulent stress and turbulence production oscillations observed in the upper layers.

The frequency of fluid ejection from the cavities to the upper layers of the flow could not be precisely determined, and, based on the movie frames, it was estimated to be around 10 vortex turns. Considering the mean tangential velocities $\overline{v}_t$ of inner-cavities vortices, the frequency of ejection is associated to wavelengths of approximately 3 $w$ and 5 $w$ for the d- and k-type roughness, respectively. These wavelengths are of the same order of the wavelengths observed in the upper layers. In this way, based on fluid ejections from cavities, we propose a ratio of approximately 10 between the wavelengths of inner-cavities vortices and that of longitudinal oscillations of upper layers.

\begin{acknowledgements}
The authors are grateful to ANP, Repsol, Cepetro, Labpetro, CNPq (grant no. 400284/2016-2), and to FAPESP (grants nos. 2012/19562-6 and 2016/13474-9) for the provided financial support.
\end{acknowledgements}


\bibliography{references}
\bibliographystyle{spphys}

\end{document}